\documentclass[a4paper,fleqn]{cas-dc}
\PassOptionsToPackage{table}{xcolor}

\usepackage[T1]{fontenc}
\usepackage{lmodern}
\usepackage{amsmath,amssymb,amsfonts}
\usepackage{bm}
\usepackage{graphicx}
\usepackage{booktabs,multirow,makecell}
\usepackage{siunitx}
\usepackage{xcolor}
\usepackage{pifont}
\usepackage{tikz}
\usetikzlibrary{arrows.meta,positioning,fit,backgrounds,calc,shapes.geometric,shapes.misc}
\usepackage{pgfplots}
\pgfplotsset{compat=1.18}
\usepgfplotslibrary{groupplots,fillbetween}
\usepackage{algorithm}
\usepackage{algpseudocode}
\usepackage[numbers,sort&compress]{natbib}
\usepackage[capitalise,nameinlink]{cleveref}  

\newtheorem{proposition}{Proposition}
\newtheorem{definition}{Definition}

\definecolor{cOurs}{HTML}{D55E00}
\definecolor{cB1}{HTML}{0072B2}
\definecolor{cB2}{HTML}{009E73}
\definecolor{cB3}{HTML}{CC79A7}
\definecolor{cB4}{HTML}{E69F00}
\definecolor{cB5}{HTML}{56B4E9}
\definecolor{cGrey}{HTML}{6B6B6B}

\pgfplotsset{
  paperstyle/.style={
    width=\columnwidth, height=0.62\columnwidth,
    tick label style={font=\scriptsize, /pgf/number format/fixed, /pgf/number format/precision=2},
    scaled ticks=false, label style={font=\footnotesize},
    legend style={font=\scriptsize, draw=none, fill=none},
    grid=major, grid style={line width=0.2pt, draw=gray!25},
    axis line style={line width=0.4pt}, tick style={line width=0.4pt},
    every axis plot/.append style={line width=0.9pt},
    cycle list={{cOurs,mark=*,mark size=1.2pt},{cB1,mark=square*,mark size=1.1pt},
                {cB2,mark=triangle*,mark size=1.4pt},{cB3,mark=diamond*,mark size=1.4pt}},
  }
}

\tikzset{
  agent/.style={draw=cB1, fill=cB1!8, rounded corners=2pt, line width=0.6pt,
                minimum width=15mm, minimum height=6mm, font=\scriptsize, align=center},
  guard/.style={draw=cOurs, fill=cOurs!8, rounded corners=2pt, line width=0.8pt,
                minimum height=6mm, font=\scriptsize\bfseries, align=center},
  plant/.style={draw=cGrey, fill=cGrey!8, rounded corners=2pt, line width=0.6pt,
                minimum width=15mm, minimum height=6mm, font=\scriptsize, align=center},
  msg/.style={-{Stealth[length=1.6mm]}, line width=0.6pt, draw=black!70},
  hal/.style={-{Stealth[length=1.6mm]}, line width=0.8pt, draw=cOurs, densely dashed},
  lbl/.style={font=\tiny, text=black!70, align=center},
}

\newcommand{\DOI}[1]{doi:~\href{https://doi.org/#1}{\nolinkurl{#1}}}
\newcommand{\best}[1]{\textbf{#1}}
\newcommand{\second}[1]{\underline{#1}}
\newcommand{\std}[1]{{\,\scriptsize$\pm$#1}}
\newcommand{\cmark}{\ding{51}}
\newcommand{\xmark}{\ding{55}}
\newproof{pf}{Proof}
\makeatletter 
\def\@opargbegintheorem#1#2#3{\trivlist
  \let\baselinestretch\@blstr
  \item[]{\bfseries #1\ #2\ (#3).}\ \itshape}
\makeatother
\pgfplotsset{paperstyle/.append style={tick label style={font=\footnotesize}, label style={font=\small},
  legend style={font=\footnotesize}, every axis plot/.append style={line width=1.1pt}}}
\newcommand{\resBusMV}{97}
\newcommand{\resBusTotal}{99}
\newcommand{\resZones}{6}
\newcommand{\resMeters}{16}
\newcommand{\resResCapMW}{45.7}
\newcommand{\resNcCapMW}{1.3}
\newcommand{\resStorageMW}{13.8}
\newcommand{\resStorageMWh}{27.5}
\newcommand{\resPeakLoadMW}{10.6}
\newcommand{\resUncontrolledStepPct}{31.0}

\newcommand{\resUncontrolledVmax}{1.080}
\newcommand{\resExpertSlackPct}{0.0}
\newcommand{\resExpertCurtPct}{8.6}

\newcommand{\resPfMaxDev}{1\times 10^{-12}}
\newcommand{\resTBusMV}{120}
\newcommand{\resTBusTotal}{122}
\newcommand{\resTResCapMW}{60.8}
\newcommand{\resTStorageMWh}{58.7}
\newcommand{\resTUncontrolledStepPct}{2.4}

\newcommand{\resTrainDays}{208}
\newcommand{\resCalDays}{39}
\newcommand{\resValDays}{39}
\newcommand{\resTestDays}{52}
\newcommand{\resClosedDays}{16}
\newcommand{\resClosedValDays}{12}

\newcommand{\resFParamsK}{22.7}
\newcommand{\resFTrainS}{53}
\newcommand{\resFRmsePct}{1.5}
\newcommand{\resEParamsK}{221}
\newcommand{\resETrainS}{57}
\newcommand{\resERmseMpu}{0.40}
\newcommand{\resDParamsM}{0.83}
\newcommand{\resDTrainMin}{82}

\newcommand{\resDParsePct}{99.9}
\newcommand{\resDBMaePct}{0.11}
\newcommand{\resDKMaePp}{0.16}
\newcommand{\resDNTrain}{36{,}000}
\newcommand{\resDEpochs}{6}

\newcommand{\resTDNTrain}{12{,}000}

\newcommand{\resPrgtParamsK}{156}
\newcommand{\resPrgtTrainMin}{9.6}

\newcommand{\resMarginMpu}{1.74}
\newcommand{\resMarginN}{4{,}124}

\newcommand{\resNatFLoadPct}{0.46}

\newcommand{\resNatDCPct}{0.36}

\newcommand{\resNatFMsgPct}{17.3}

\newcommand{\resNatDMsgPct}{0.38}

\newcommand{\resNatDPropSharePct}{36.8}

\newcommand{\resSeeds}{3}
\newcommand{\resSeedsClosed}{3}
\newcommand{\resAuprcOurs}{0.846}

\newcommand{\resTprOursPct}{81.4}
\newcommand{\resBestBaselineName}{B4 Autoencoder}
\newcommand{\resBestBaselineShort}{B4}
\newcommand{\resAuprcBestBaseline}{0.837}
\newcommand{\resTprBestBaselinePct}{77.4}

\newcommand{\resAuprcOursCiLo}{0.815}
\newcommand{\resAuprcOursCiHi}{0.879}
\newcommand{\resAuprcBestCiLo}{0.814}
\newcommand{\resAuprcBestCiHi}{0.858}
\newcommand{\resAuprcDiffCiLo}{-0.014}
\newcommand{\resAuprcDiffCiHi}{0.032}

\newcommand{\resWilcoxonPairs}{52}

\newcommand{\resWilcoxonNSigBase}{4}

\newcommand{\resAuprcBfourTol}{0.811}

\newcommand{\resDeadbandBfourTol}{0.25}

\newcommand{\resWilcoxonBfiveTolP}{0.106}
\newcommand{\resAuprcBfiveTol}{0.874}

\newcommand{\resAuprcEBfiveTol}{0.826}
\newcommand{\resAuprcDBfiveTol}{0.705}

\newcommand{\resTprBfiveTolPct}{86.2}
\newcommand{\resDeadbandBfiveTol}{1.00}
\newcommand{\resAurocBfiveTol}{0.942}

\newcommand{\resAuprcBfiveTolCiLo}{0.852}
\newcommand{\resAuprcBfiveTolCiHi}{0.898}
\newcommand{\resAuprcDiffBfiveTolCiLo}{-0.060}
\newcommand{\resAuprcDiffBfiveTolCiHi}{0.005}

\newcommand{\resAuprcAllOurs}{0.457}
\newcommand{\resAuprcAllBest}{0.489}
\newcommand{\resAuprcAllBestName}{B4}
\newcommand{\resMsgPrevalenceAllPct}{9.1}

\newcommand{\resMsgPrevalencePct}{3.7}

\newcommand{\resAuprcFOurs}{0.856}

\newcommand{\resAuprcEOurs}{0.967}

\newcommand{\resAuprcDOurs}{0.905}
\newcommand{\resAuprcDBest}{0.739}
\newcommand{\resAuprcDBestName}{B1}

\newcommand{\resFarOursOnePct}{1.17}

\newcommand{\resFarOursFivePct}{5.07}
\newcommand{\resFarAllOursFivePct}{7.42}
\newcommand{\resFarEpiOursFivePct}{16.74}
\newcommand{\resFarFixedFivePct}{0.00}
\newcommand{\resFarChiFivePct}{0.9}
\newcommand{\resFarOursTenPct}{9.96}

\newcommand{\resFarMaxGapPp}{0.17}
\newcommand{\resFdrOurs}{0.181}
\newcommand{\resFdrFreeOurs}{0.107}
\newcommand{\resFdrEpiOurs}{0.468}

\newcommand{\resDelayOurs}{0.01}
\newcommand{\resDelayMissedOursPct}{2.1}
\newcommand{\resDelayMissedBestPct}{6.2}

\newcommand{\resTypeHFourOurs}{0.224}

\newcommand{\resTypeHSevenOurs}{0.438}

\newcommand{\resTypeHEightOurs}{0.577}

\newcommand{\resRootAccOursPct}{82.2}
\newcommand{\resRootAccDetOursPct}{94.7}
\newcommand{\resRootAccTopoPct}{64.5}
\newcommand{\resRootAccBestPct}{78.3}
\newcommand{\resRootAccBestName}{B5}
\newcommand{\resTopAKey}{B4}
\newcommand{\resTopAName}{B4 Autoencoder}
\newcommand{\resTopAColor}{cB4}
\newcommand{\resTopBKey}{B5}
\newcommand{\resTopBName}{B5 Self-consist.}
\newcommand{\resTopBColor}{cB5}
\newcommand{\resTopCKey}{B1}
\newcommand{\resTopCName}{B1 Rules}
\newcommand{\resTopCColor}{cB1}
\newcommand{\resSweepTprOursLowPct}{1.6}
\newcommand{\resSweepTprOursHighPct}{27.0}
\newcommand{\resSweepTprTopLowPct}{29.5}
\newcommand{\resSweepTprTopHighPct}{100.0}

\newcommand{\resSweepCphShareHighPct}{26.9}
\newcommand{\resSweepMax}{8}
\newcommand{\resViolUnguarded}{0.0700}

\newcommand{\resCurtOurs}{32.7}

\newcommand{\resCurtExtraPct}{8.9}

\newcommand{\resExecUnguarded}{37.7}

\newcommand{\resViolBone}{0.0022}

\newcommand{\resViolBtwo}{0.0002}

\newcommand{\resExecBthree}{9.7}

\newcommand{\resExecBfour}{6.3}
\newcommand{\resViolBfive}{0.0009}

\newcommand{\resViolBsix}{0.0002}

\newcommand{\resCurtBseven}{50.0}

\newcommand{\resCurtExtraMinDetPct}{7.3}
\newcommand{\resCurtExtraMaxDetPct}{33.4}

\newcommand{\resEpisodesClosed}{126}
\newcommand{\resRepairPct}{6.4}
\newcommand{\resRequeryPct}{18.9}

\newcommand{\resPpDevMax}{6\times 10^{-11}}
\newcommand{\resClosedSteps}{1{,}536}
\newcommand{\resViolWilcoxonUnguardedP}{0.0046}

\newcommand{\resClosedStatDays}{16}

\newcommand{\resViolWilcoxonMinBaseP}{0.166}
\newcommand{\resViolWilcoxonMinBaseName}{B4}

\newcommand{\resRepCovOursPct}{95.6}

\newcommand{\resRepCovFeOursPct}{84.0}

\newcommand{\resRepNOurs}{293}

\newcommand{\resRepCovMinDetPct}{44.5}
\newcommand{\resRepCovMaxDetPct}{96.8}
\newcommand{\resRepCovNominalPct}{95}
\newcommand{\resRepErrMedianMpu}{2.74}
\newcommand{\resRepEstMedianMpu}{2.77}
\newcommand{\resRepLinMedianMpu}{0.05}
\newcommand{\resMarginClMpu}{6.69}
\newcommand{\resMarginClN}{319}

\newcommand{\resMarginClPersistMpu}{32.40}

\newcommand{\resClosedCalDays}{12}
\newcommand{\resRepCovOfflinePct}{33.2}

\newcommand{\resViolUnguardedCiLo}{0.0418}
\newcommand{\resViolUnguardedCiHi}{0.1038}
\newcommand{\resIncidentAgent}{D}
\newcommand{\resIncidentType}{H2}
\newcommand{\resIncidentTypeName}{unit or scale slip}

\newcommand{\resIncidentVUnguarded}{1.067}
\newcommand{\resIncidentVOurs}{1.043}
\newcommand{\resIncidentBuses}{7}

\newcommand{\resClUpEpisodes}{257}
\newcommand{\resClUpReached}{3}
\newcommand{\resCaseDay}{25}
\newcommand{\resCaseBaselineName}{B7 Safety filter}
\newcommand{\resCaseBaselineColor}{cB4}
\newcommand{\resCaseVmaxUnguarded}{1.066}
\newcommand{\resCaseVmaxOurs}{1.048}
\newcommand{\resCaseVmaxBaseline}{1.046}
\newcommand{\resCaseEpisodes}{18}
\newcommand{\resAuprcNoRes}{0.773}
\newcommand{\resAuprcNoBias}{0.868}
\newcommand{\resAuprcNoSynth}{0.363}
\newcommand{\resRootHeadNoSynth}{0.295}
\newcommand{\resRootTopoNoSynth}{0.307}

\newcommand{\resAuprcSingle}{0.865}
\newcommand{\resFdrNoBh}{0.984}
\newcommand{\resFarSingleFivePct}{5.00}
\newcommand{\resFarClassMinSinglePct}{0.6}
\newcommand{\resFarClassMaxSinglePct}{26.0}
\newcommand{\resFarClassMinOursPct}{4.8}
\newcommand{\resFarClassMaxOursPct}{7.7}

\newcommand{\resAuprcFullAbl}{0.875}
\newcommand{\resRootHeadAbl}{0.809}
\newcommand{\resRootTopoAbl}{0.592}

\newcommand{\resCurtFullAbl}{33.6}
\newcommand{\resCurtAltGround}{32.2}
\newcommand{\resCurtNoRequery}{33.4}

\newcommand{\resViolNoMargin}{0.0141}

\newcommand{\resTransferAuprcIn}{0.788}
\newcommand{\resTransferAuprcZero}{0.719}

\newcommand{\resTransferFarRecalPct}{5.01}
\newcommand{\resTransferFarNoRecalPct}{5.97}

\newcommand{\resTransferBestBaseName}{B5 Self-consist.}

\newcommand{\resTransferAuprcBestBase}{0.783}

\newcommand{\resTransferNBaseAboveZero}{2}

\newcommand{\resLotoMedianDrop}{0.016}

\newcommand{\resLotoNSmallDrop}{6}
\newcommand{\resTypeHSevenMin}{0.116}
\newcommand{\resTypeHSevenMax}{0.902}

\newcommand{\resLotoSeenHFour}{0.174}
\newcommand{\resLotoUnseenHFour}{0.043}

\newcommand{\resLotoSeenHSeven}{0.902}
\newcommand{\resLotoUnseenHSeven}{0.237}

\newcommand{\resMissingAuprcMax}{0.847}
\newcommand{\resMissingFarMaxPct}{4.91}

\newcommand{\resLatMedianMain}{9.3}
\newcommand{\resLatPninetyfiveMain}{10.0}
\newcommand{\resLatMaxFeeder}{18.7}

\newcommand{\resLatOurs}{4.0}
\newcommand{\resLatBfive}{99.7}
\newcommand{\resLatStepShare}{0.0011}
\newcommand{\resAuprcFBfour}{0.889}

\newcommand{\resFpAlteredBfourPct}{95}

\newcommand{\resAuprcFBfive}{0.986}

\newcommand{\resFpAlteredBfivePct}{100}

\newcommand{\resTypeMinBfive}{0.931}
\newcommand{\resAuprcMinAgentOurs}{0.856}
\newcommand{\resAuprcMinAgentBest}{0.689}

\newcommand{\resPrgtEpochsMax}{10}
\newcommand{\resSubDimLoad}{10}
\newcommand{\resSubDimRes}{5}
\newcommand{\resSubDimVolt}{9}
\newcommand{\resPrgtEpochSteps}{4{,}800}
\newcommand{\resFpAlteredPct}{48}

\newcommand{\resNegAlteredPct}{7.6}

\newcommand{\resFdrRejectingOurs}{0.85}
\newcommand{\resNatRecallDCOursPct}{100.0}

\newcommand{\resNatRecallFLoadOursPct}{26.9}
\newcommand{\resNatRecallFLoadBfivePct}{19.4}

\newcommand{\resNatCountDC}{18}
\newcommand{\resValAuprcBase}{0.764}
\newcommand{\resValAuprcFrozen}{0.851}

\newcommand{\resValBestName}{B5}
\newcommand{\resValMargin}{+0.008}

\newcommand{\resLoopTried}{8}
\newcommand{\resLoopKept}{4}
\newcommand{\resValViolUnguarded}{0.3401}

\newcommand{\resValViolRedPct}{97.4}
\newcommand{\resValCurtExtraPct}{9.5}
\newcommand{\resValViolRec}{0.0439}
\newcommand{\resValViolHybrid}{0.0089}

\newcommand{\resValCurtPersist}{1434}
\newcommand{\resValCurtExpert}{158}

\begin{document}
\let\WriteBookmarks\relax
\def\floatpagepagefraction{1}
\def\textpagefraction{.001}

\shorttitle{GridAnchor: Safeguarding Autonomous Energy Systems Against Cyber-Physical Hallucinations}
\shortauthors{A. Bahi and H. Masrur}

\title[mode=title]{GridAnchor: Specialized Multi-Agent AI for Safeguarding Autonomous Energy Systems Against Cyber-Physical Hallucinations}

\author[1]{Abderaouf Bahi}[orcid=0009-0003-7116-5080]
\cormark[1]
\ead{a.bahi@univ-eltarf.dz}
\credit{Writing -- original draft, Software, Validation, Methodology, Investigation, Conceptualization}

\affiliation[1]{organization={Computer Science and Applied Mathematics Laboratory (LIMA), Faculty of Science and Technology, Chadli Bendjedid University},
            addressline={P.O. Box 73},
            city={El Tarf},
            postcode={36000},
            country={Algeria}}

\author[2,3]{Hasan Masrur}[orcid=0000-0002-6695-2851]
\ead{hasan.masrur@kfupm.edu.sa}
\credit{Writing -- review and editing, Resources, Supervision}

\affiliation[2]{organization={Center for Smart Mobility and Logistics (IRC-SML), King Fahd University of Petroleum \& Minerals},
            city={Dhahran},
            postcode={31261},
            country={Saudi Arabia}}

\affiliation[3]{organization={Department of Industrial and Systems Engineering, King Fahd University of Petroleum \& Minerals},
            city={Dhahran},
            postcode={31261},
            country={Saudi Arabia}}

\cortext[1]{Corresponding author}

\begin{abstract}
Autonomous energy systems increasingly chain specialized learned agents, such as a forecaster, a state estimator and a dispatcher that may be a language model.
These agents can emit confident outputs that contradict physics, and downstream agents may act on them.
We define such cyber-physical hallucinations and their cascades, organize them in eight types, and build CPH-Bench, a generator of labeled episodes on SimBench feeders.
We propose GridAnchor, a verification layer whose Physics-Residual Graph Transformer scores every claim from residuals between the claims, the measurements and a linearized grid model.
Mondrian conformal p-values with the Benjamini-Hochberg procedure turn the scores into decisions, a root head attributes cascades, and a quadratic program repairs unsafe actions with a voltage margin calibrated at repaired steps.
On a simulated rural medium-voltage feeder, GridAnchor reaches a message-level area under the precision-recall curve of \resAuprcOurs, against \resAuprcBestBaseline{} for the strongest of six baselines, a difference within the confidence interval.
It leads on estimates and actions, whereas a self-consistency check that knows the label tolerances reaches \resAuprcBfiveTol{} by leading on forecasts.
The false-alarm rate stays within \resFarMaxGapPp{} percentage points of the nominal level on steps without injected hallucinations.
In closed loop, GridAnchor leaves no bus outside the voltage band on the test days, against \resViolUnguarded\% of the bus-steps without protection, and simpler detectors with the same mitigation come close.
Stale but coherent messages and single-bus forecast errors remain mostly undetected. Project details are available at \url{https://github.com/AraoufBh/GridAnchor}
\end{abstract}


\maketitle

\section{Introduction}
\label{sec:intro}
Autonomous energy systems coordinate large numbers of distributed energy resources with little human supervision \cite{kroposki2020autonomous}.
Their control is moving from single controllers to teams of specialized AI agents that forecast (F), estimate the state (E) and dispatch flexible resources (D) \cite{mcarthur2007multi,chen2022reinforcement,cao2021data}.
Language-model agents \cite{wang2024survey} are entering power-system workflows as well \cite{majumder2024exploring,yan2024realtime,jia2025enhancing}, and small specialized language models have been proposed as the practical choice for repetitive agentic tasks \cite{belcak2025small}.
In such a team, each agent consumes the outputs of the others as if they were measurements.

These outputs are numeric claims that carry no statement of uncertainty.
A learned agent can emit a confident and well-formed claim that is physically impossible, that contradicts the measurements, or that misdescribes the consequence of its own action.
We call such an output a cyber-physical hallucination (CPH).
The term is borrowed from language models, whose hallucinations are fluent statements unsupported by facts \cite{huang2025survey}; here the facts are the physical state of the grid.
A CPH emitted by one agent is consumed by the next ones, and the resulting cascade can end in an unsafe physical action.

In our closed-loop simulations, hallucinations of F and E propagate into the messages of D, but only \resClUpReached{} of the \resClUpEpisodes{} episodes injected into F and E led to an unsafe executed action.
Most unsafe actions came from hallucinations of D itself.
In the worst case, a hallucination of type \resIncidentType{} (\resIncidentTypeName) in the message of agent \resIncidentAgent{} raised the highest bus voltage to \resIncidentVUnguarded~p.u., above the limit of 1.05~p.u. at \resIncidentBuses{} buses.
No measurement was faulty and no agent reported an error.
With the verifier proposed in this paper, the same step ended at \resIncidentVOurs~p.u.

Existing tools each cover a fragment of this problem.
Bad-data detection and the detection of false data injection assume errors in measurements, with known noise models \cite{abur2004power,liu2011false,musleh2020survey}.
Hallucination detectors for language models rely on self-consistency or on semantic uncertainty and ignore physics \cite{manakul2023selfcheckgpt,farquhar2024detecting,huang2025survey}.
Safety filters correct a proposed action but ignore upstream cascades and attribution \cite{wabersich2021predictive}.
Failure taxonomies of multi-agent language systems \cite{cemri2025why} and external-verifier frameworks \cite{kambhampati2024llms} lack physical grounding and statistical guarantees.

\begin{figure*}[t]
\centering
\begin{tikzpicture}[x=1mm, y=1mm, font=\footnotesize,
    agent/.append style={font=\footnotesize, minimum width=17mm, minimum height=8mm},
    guard/.append style={font=\footnotesize\bfseries, minimum height=7mm},
    lbl/.append style={font=\scriptsize},
    feeder/.style={plant, minimum width=18mm, minimum height=17mm}]
  \newcommand{\minivolt}[3]{%
    \begin{scope}[shift={(#1)}]
      \node[font=\footnotesize] at (0,5.6) {Feeder};
      \draw[line width=0.35pt, draw=black!60] (-7,-7) -- (7,-7);
      \draw[line width=0.35pt, draw=black!60] (-7,-7) -- (-7,2.6);
      \draw[densely dotted, line width=0.5pt] (-7,-1.2) -- (7,-1.2);
      \node[font=\scriptsize, anchor=south west, inner sep=0.6pt] at (-6.6,-1.1) {$v_{\max}$};
      \draw[line width=0.8pt, draw=#2] #3;
    \end{scope}}
  \begin{scope}
    \node[agent] (F) at (0,0) {Forecast\\agent F};
    \node[agent] (E) at (21,0) {Estimation\\agent E};
    \node[agent] (D) at (42,0) {Dispatch\\agent D};
    \node[feeder] (G) at (66,-4) {};
    \minivolt{66,-4}{cOurs}{(-6.5,-4.6) .. controls (-3,-4.4) and (-1.5,-3.6) .. (0,0.9) .. controls (1.2,2.4) and (3.5,1.6) .. (6.5,0.9)}
    \draw[msg] (F) -- (E);
    \draw[msg] (E) -- (D);
    \draw[msg] (D.east) -- (D.east -| G.west);
    \draw[msg, draw=cGrey] (G.south) -- ++(0,-6) -| (E.south);
    \node[lbl] at (44,-20.8) {measurements $\bm{y}_t$};
    \draw[hal] (F.north) to[bend left=40] node[lbl, above, text=cOurs] {forecast $\times 10$} (E.north);
    \draw[hal] (E.north) to[bend left=40] node[lbl, above, text=cOurs] {biased voltages} (D.north);
    \draw[hal] (D.north) to[bend left=30] node[lbl, above, text=cOurs] {under-curtailment} ([xshift=-3mm]G.north);
    \node[lbl, text=cOurs, anchor=east] at (64.5,-15.2) {overvoltage};
    \node[font=\footnotesize] at (33,-34) {(a) Unguarded multi-agent pipeline};
  \end{scope}
  \begin{scope}[shift={(92,0)}]
    \node[agent] (F) at (0,0) {Forecast\\agent F};
    \node[agent] (E) at (21,0) {Estimation\\agent E};
    \node[agent] (D) at (42,0) {Dispatch\\agent D};
    \node[feeder] (G) at (66,-4) {};
    \minivolt{66,-4}{cB2}{(-6.5,-4.6) .. controls (-3,-4.4) and (-1.5,-3.9) .. (0,-3.0) .. controls (1.2,-2.6) and (3.5,-2.9) .. (6.5,-3.0)}
    \node[guard, minimum width=57mm] (V) at (21,-17) {GridAnchor: verify, attribute, ground, repair};
    \draw[hal] (F.south) -- (F.south |- V.north) node[lbl, pos=0.5, left, text=cOurs] {flagged\\claim};
    \draw[msg] ([xshift=-2.5mm]E.south) -- ([xshift=-2.5mm]E.south |- V.north);
    \draw[msg] ([xshift=-2.5mm]D.south) -- ([xshift=-2.5mm]D.south |- V.north);
    \draw[msg, draw=cB2] ([xshift=2.5mm]E.south |- V.north) -- ([xshift=2.5mm]E.south) node[lbl, pos=0.5, right, text=cB2] {grounded};
    \draw[msg, draw=cB2] ([xshift=2.5mm]D.south |- V.north) -- ([xshift=2.5mm]D.south) node[lbl, pos=0.5, right, text=cB2] {re-query};
    \draw[msg, draw=cB2] (V.east) -| ([xshift=-5mm]G.south);
    \node[lbl, text=cB2] at (56.2,-21.2) {verified or\\repaired $\bm{u}_t$};
    \draw[msg, draw=cGrey] ([xshift=5mm]G.south) -- ++(0,-14.5) -| (V.south);
    \node[lbl] at (47,-29.2) {measurements $\bm{y}_t$};
    \node[font=\footnotesize] at (33,-34) {(b) Pipeline guarded by GridAnchor};
  \end{scope}
\end{tikzpicture}
\caption{A hallucination cascade (a) without and (b) with GridAnchor (schematic).}
\label{fig:concept}
\end{figure*}
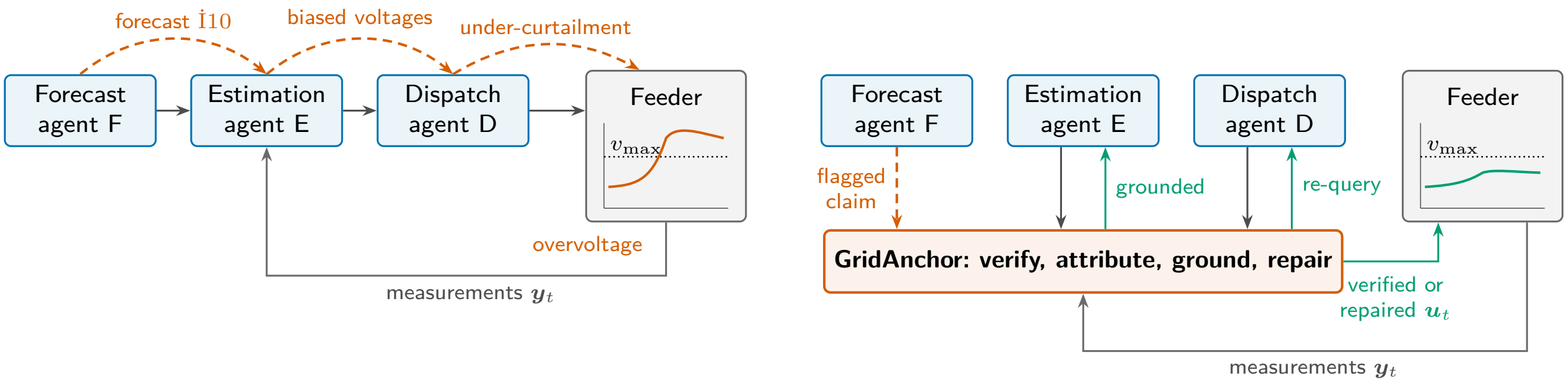

We study how often and in which forms specialized learned agents produce CPHs in a distribution grid (RQ1).
We ask whether a physics-grounded learned verifier detects them better than generic detectors at a controlled false-alarm rate (RQ2) and attributes a cascade to its root agent (RQ3).
We test whether detection, grounded re-query and repair reduce operational violations without a large curtailment penalty (RQ4).
Finally, we check whether the verifier transfers to an unseen feeder and to unseen types fast enough for steps of 15~min (RQ5).
The answers come from GridAnchor, a verification layer placed between the agents and the plant (\cref{fig:concept}), and from four contributions.
\begin{itemize}
\item Formalization and benchmark. We define CPHs and their cascades, organize them in a taxonomy of eight types, and build CPH-Bench, a generator of labeled hallucination episodes on the SimBench grids and profiles \cite{meinecke2020simbench}.
\item Verifier. The Physics-Residual Graph Transformer (PRGT) reads residual tokens computed from the claims, the measurements and a linearized grid model, and is trained by taxonomy-driven synthesis to detect, reconstruct and attribute claims.
\item Guarantees. Mondrian conformal p-values with the Benjamini-Hochberg procedure control false alarms and false discoveries under stated exchangeability assumptions, and a voltage margin calibrated at repaired steps makes a minimum-norm repair feasible with a stated probability.
\item Evidence. Open-loop and closed-loop experiments compare GridAnchor with seven baselines and two tolerance-aware variants, with ablations and studies of calibration, severity, transfer, robustness and latency.
\end{itemize}

On the main feeder, GridAnchor reaches a message-level area under the precision-recall curve of \resAuprcOurs, against \resAuprcBestBaseline{} for the strongest of six detection baselines, a difference that the test days do not resolve.
In closed loop it leaves no bus outside the voltage band on the test days, with \resCurtExtraPct\% more curtailment than an expert that knows the true state.
The study also reports negative findings.
A self-consistency check that knows the label tolerances detects better overall, simpler detectors with the same mitigation come close in closed loop, and refitted baselines outperform the transferred verifier.

The remainder of the paper is organized as follows.
\Cref{sec:related} reviews related work, \cref{sec:problem} formalizes the problem, \cref{sec:method} presents GridAnchor, \cref{sec:setup} describes the experimental setup, \cref{sec:results} reports the results, \cref{sec:discussion} discusses them and \cref{sec:conclusion} concludes.
Appendix~\ref{sec:proofs} contains the proofs, and Appendix~\ref{sec:hyper} lists the hyperparameters and the grammar of the dispatch agent.

\section{Related Work}
\label{sec:related}

\subsection{AI and Multi-Agent Control in Autonomous Energy Systems}
Multi-agent systems have a long history in power engineering \cite{mcarthur2007multi}, and reinforcement learning is studied for voltage control, frequency regulation and energy management, with safety among its open challenges \cite{chen2022reinforcement}.
Multi-agent deep reinforcement learning has been applied to decentralized voltage control in feeders with high photovoltaic penetration \cite{cao2021data}.
These works train the agents to act well; they do not verify what the agents tell each other at run time.

\subsection{Language-Model Agents in Power Systems}
Language-model agents plan, call tools and cooperate in teams \cite{wang2024survey}, and their use in the energy sector comes with a warning about reliability \cite{majumder2024exploring}.
A GPT-based agent has been coupled with deep reinforcement learning to enforce linguistic stipulations in real-time optimal power flow \cite{yan2024realtime}.
A feedback-driven multi-agent framework has raised the success rate of language models in power-system simulation tasks \cite{jia2025enhancing}.
Small language models have been argued to suit narrow and repetitive agentic tasks better than general models \cite{belcak2025small}, the view that our dispatch agent follows.

\subsection{Hallucination Detection}
The hallucinations of large language models (LLMs) have been classified by cause and by type, together with detection and mitigation methods \cite{huang2025survey}.
SelfCheckGPT flags content that is inconsistent across sampled answers \cite{manakul2023selfcheckgpt}, and semantic entropy measures the uncertainty of their meaning \cite{farquhar2024detecting}; both need repeated sampling and ignore physical quantities.
Failure analyses of multi-agent language systems have produced taxonomies in which inter-agent misalignment is a major category \cite{cemri2025why}, and the LLM-Modulo proposal pairs a language model with external verifiers that critique its plans \cite{kambhampati2024llms}.
Neither provides a statistical guarantee on the detection, and neither is tied to a physical model.

\subsection{Physics-Grounded Safety, Bad Data and Attack Detection}
Power-system state estimation detects bad data with the chi-square test and the largest normalized residual test under a Gaussian noise model \cite{abur2004power}.
False data injection attacks can be built to bypass these residual tests \cite{liu2011false}, which has motivated a large body of model-based and data-driven detectors \cite{musleh2020survey}.
These methods target corrupted measurements, whereas a hallucinating agent corrupts claims that were computed from correct measurements.
Predictive safety filters modify the action of a learning-based controller as little as possible so that constraints hold \cite{wabersich2021predictive}, but they act on the last agent only and trust the state and forecasts they receive.

\subsection{Conformal Inference for Anomaly Detection}
Conformal prediction turns any score into a statistically valid statement under exchangeability \cite{vovk2022algorithmic}.
Its split variant needs only a held-out calibration set \cite{angelopoulos2023conformal}.
Conformal p-values for outlier detection are positively dependent, so that the BH procedure \cite{benjamini1995controlling} controls the false discovery rate of many tests \cite{bates2023testing}; we apply this machinery to agent claims with class-conditional calibration.

Few studies verify the messages that learned agents exchange in an energy system, so \cref{tab:positioning} compares families of methods rather than individual systems.
The six properties are physical grounding (PG), awareness of multi-agent cascades (CA), a learned detector (LD), a finite-sample guarantee (FG), root-cause attribution (RA) and action repair (AR).
No earlier family is at once physically grounded, aware of cascades, equipped with a finite-sample guarantee and able to repair an action.
GridAnchor combines the six properties for chains of specialized agents in distribution grids.

\begin{table}[t]
\centering
\caption{Properties of related approaches.}
\label{tab:positioning}
\setlength{\tabcolsep}{2.8pt}
\begin{tabular}{@{}>{\raggedright\arraybackslash}p{46mm}cccccc@{}}
\toprule
Line of work & PG & CA & LD & FG & RA & AR \\
\midrule
Bad-data tests \cite{abur2004power} & \cmark & \xmark & \xmark & \xmark & \xmark & \xmark \\
False data injection detection \cite{liu2011false,musleh2020survey} & \cmark & \xmark & \cmark & \xmark & \xmark & \xmark \\
LLM hallucination detection \cite{manakul2023selfcheckgpt,farquhar2024detecting,huang2025survey} & \xmark & \xmark & \cmark & \xmark & \xmark & \xmark \\
Multi-agent failures, verifiers \cite{cemri2025why,kambhampati2024llms} & \xmark & \cmark & \xmark & \xmark & \xmark & \xmark \\
LLM agents with power tools \cite{yan2024realtime,jia2025enhancing} & \cmark & \xmark & \xmark & \xmark & \xmark & \xmark \\
Safety filters, safe learning \cite{wabersich2021predictive,chen2022reinforcement} & \cmark & \xmark & \xmark & \xmark & \xmark & \cmark \\
Conformal outlier tests \cite{bates2023testing,angelopoulos2023conformal} & \xmark & \xmark & \cmark & \cmark & \xmark & \xmark \\
\midrule
GridAnchor (this work) & \cmark & \cmark & \cmark & \cmark & \cmark & \cmark \\
\bottomrule
\end{tabular}

\end{table}

\section{Problem Formulation}
\label{sec:problem}

\subsection{System Model}
We consider a distribution feeder with $N$ monitored buses, operated in steps of $\Delta t = 15$~min indexed by $t$.
Bus $n$, in zone $z(n)$ of $Z$ zones, hosts a load $p^{\mathrm{L}}_{n,t}$, curtailable renewable generation with available power $p^{\mathrm{R}}_{n,t}$, telemetered non-controllable generation $p^{\mathrm{G}}_{n,t}$ and possibly a storage unit.
The action $\bm{u}_t = (\bm{b}_t, \bm{\kappa}_t) \in \mathbb{R}^{2Z}$ sets the storage power $b_{z,t}$ of each zone (positive when charging), shared among its units with shares $\sigma_n$ proportional to their ratings, and the curtailment fraction $\kappa_{z,t} \in [0,1]$ of its renewable generation.
The net active injection at bus $n$ is
\begin{equation}
p_{n,t} = -p^{\mathrm{L}}_{n,t} + (1-\kappa_{z(n),t})\,p^{\mathrm{R}}_{n,t} + p^{\mathrm{G}}_{n,t} - \sigma_n b_{z(n),t},
\label{eq:injection}
\end{equation}
and the voltage magnitudes $\bm{v}_t \in \mathbb{R}^N$ follow from the AC power flow equations.
The state of charge $e_{z,t} \in [0,1]$ of zone $z$ evolves with the charging and discharging efficiency $\eta$ and must stay within $[e_{\min}, e_{\max}]$, while voltages must stay within the band $[v_{\min}, v_{\max}]$.

Three specialized agents form a directed chain from data to action.
The forecasting agent F claims the load and the available renewable power of every bus for step $t$.
The state-estimation agent E reads the measurements $\bm{y}_t$ and the message of F, and claims the voltage of every bus at the start of the step, under the previous action $\bm{u}_{t-1}$.
The dispatch agent D reads the messages of F and E and the measured state of charge, and emits $\bm{u}_t$ with two rationale values.
These are the maximum voltage $X$ that it expects after the action and the total curtailed power $C$.
The measurements $\bm{y}_t$ are the active power, reactive power and voltage at the substation, the voltage at the metered buses $\mathcal{M}$, the state of charge of each zone and the non-controllable generation.
None of the messages carries an uncertainty estimate, the usual situation when point predictors and token-generating models are chained.

\subsection{Claims and Cyber-Physical Hallucinations}
\begin{definition}[Claim]
A \emph{claim} is a tuple $c = (a, k, \ell, t, v)$ with emitting agent $a \in \{\mathrm{F}, \mathrm{E}, \mathrm{D}\}$, quantity type $k$, location $\ell$ (a bus, a zone or the system), time step $t$ and asserted value $v$, transmitted without an uncertainty estimate.
A \emph{message} is the set of claims that one agent emits at one step.
\end{definition}

\begin{definition}[Cyber-physical hallucination]
\label{def:cph}
Let $v^\star$ be the physically realized value of the claimed quantity.
A value claim is a \emph{cyber-physical hallucination} (CPH) if $|v - v^\star| > \epsilon_k$, where $\epsilon_k$ is an operational tolerance: $0.01$~p.u. for voltages, $10\%$ of the device rating for active power, $5$ percentage points for the state of charge, $0.01$~p.u. for a stated post-action voltage and $10\%$ of the mean zone renewable capacity for the stated curtailment.
An action claim is a CPH if (i) it violates a device limit beyond the tolerance, or (ii) its execution leaves a bus voltage outside the band by more than $0.002$~p.u. although the expert controller finds an action without voltage slack in the same state, or (iii) the rationale claims of the same message contradict the physical consequence of the action by more than $\epsilon_k$, or (iv) the message cannot be parsed.
A CPH is a \emph{root} CPH if it originates in the generation of its own agent, and a \emph{propagated} CPH if it vanishes when the upstream inputs of that agent are replaced by their ground-truth values.
\end{definition}

\begin{definition}[Cascade]
The \emph{cascade} of a root CPH is the set of propagated CPHs that it causes in downstream agents at the same or later steps.
Its depth is the number of agents on the longest causal path, and it \emph{reaches the plant} if it produces an executed action that is a CPH.
\end{definition}

Measurement noise and ordinary forecast error stay within the tolerance and are not hallucinations.
A cyber attack such as false data injection \cite{liu2011false} targets the measurements from outside, whereas a CPH is produced by an agent from correct or already corrupted inputs.
Case (iii) covers the unfaithful explanation and case (iv) the format failures of token-generating agents.
Because the tolerances are operational choices, the natural hallucination rates are also reported with all tolerances scaled by $0.5$ and by $2$.

\subsection{Taxonomy}
\Cref{tab:taxonomy} lists the eight CPH types used throughout the paper, with their analogues in the language-model literature \cite{huang2025survey}.
Types H1 to H6 corrupt individual claims, H7 adds a slow ramp expressed in units of the clean error spread of the claim, and H8 makes the rationale of D disagree with its action.
The taxonomy drives both the synthesis of training data for the verifier and the injection protocol of the benchmark.
A verifier of such hallucinations must flag claims at a false-alarm probability fixed in advance and name the root agent.
It must return an action that is feasible with a stated probability and stays close to the proposal of D.
It must run in a small fraction of the 15~min step, without any learned parameter that depends on the number of buses.

\begin{table}[t]
\centering
\caption{Taxonomy of cyber-physical hallucinations.}
\label{tab:taxonomy}
\setlength{\tabcolsep}{2.4pt}
\scriptsize
\begin{tabular}{@{}p{4mm}p{29mm}p{15mm}p{7mm}p{25mm}@{}}
\toprule
 & Definition & Language-model analogue & Agents & Severity (low, medium, high) \\
\midrule
H1 & Confabulated value: claim replaced by a draw from its training marginal & fabricated fact & F, E, D & 1, 5, 20 buses or 1, 2, 3 zones \\
H2 & Unit or scale slip: claim multiplied by 10, 0.1 or 1000 & unit or magnitude error & F, D & 1, 5, 20 buses or 1, 2, 3 zones \\
H3 & Sign inversion of storage power or of the voltage deviation from 1 p.u. & negation error & E, D & 1, 5, 20 buses or 1, 2, 3 zones \\
H4 & Stale replay: value of the same claim 4 or 96 steps earlier & outdated knowledge & F, E, D & one zone at lag 4; whole message at lag 4; at lag 96 \\
H5 & Entity swap: values of two buses or two zones exchanged & entity confusion & F, E, D & 1, 3, 8 bus pairs or 1, 1, 2 zone pairs \\
H6 & Physically impossible: output above the envelope, negative load, voltage above 1.15 p.u., power beyond the rating & logical contradiction & F, E, D & 1, 5, 20 buses or 1, 2, 3 zones \\
H7 & Subtle drift: ramp reaching $s$ times the clean error spread of the claim, on one zone & gradual context drift & F, E & $s = 2, 4, 8$ \\
H8 & Unfaithful rationale: $X$ or $C$ contradicts the action & unfaithful explanation & D & growing mismatch \\
\bottomrule
\end{tabular}

\end{table}

\section{GridAnchor}
\label{sec:method}

\subsection{Overview}
\begin{figure*}[t]
\centering
\resizebox{\textwidth}{!}{
\begin{tikzpicture}[x=1mm, y=1mm, font=\footnotesize,
    lbl/.append style={font=\scriptsize},
    blk/.style={draw=black!70, fill=black!3, rounded corners=2pt, line width=0.6pt, align=center, minimum height=22mm, inner sep=1.5pt},
    bus/.style={circle, draw=cB1, fill=cB1!15, line width=0.5pt, inner sep=0pt, minimum size=2.1mm},
    busflag/.style={circle, draw=cOurs, fill=cOurs!35, line width=0.6pt, inner sep=0pt, minimum size=2.1mm},
    zonetok/.style={regular polygon, regular polygon sides=6, draw=cB2, fill=cB2!15, line width=0.5pt, inner sep=0pt, minimum size=3.2mm},
    systok/.style={rectangle, draw=cGrey, fill=cGrey!20, line width=0.5pt, inner sep=0pt, minimum size=2.6mm},
    layer/.style={draw=cB1, fill=cB1!8, line width=0.5pt, minimum width=16mm, minimum height=4.2mm, inner sep=0pt, font=\scriptsize},
    head/.style={draw=cB1, fill=white, rounded corners=1pt, line width=0.5pt, minimum width=21mm, minimum height=5.6mm, inner sep=1pt, font=\scriptsize, align=center}]
  \node[agent, minimum width=13mm, minimum height=4.6mm] (mf) at (7,8.4) {$M^{\mathrm{F}}_t$};
  \node[agent, minimum width=13mm, minimum height=4.6mm] (me) at (7,2.8) {$M^{\mathrm{E}}_t$};
  \node[agent, minimum width=13mm, minimum height=4.6mm] (md) at (7,-2.8) {$M^{\mathrm{D}}_t$};
  \node[plant, minimum width=13mm, minimum height=4.6mm] (yy) at (7,-8.4) {$\bm{y}_t$};
  \node[lbl] at (7,-14.2) {messages and\\measurements};
  \node[blk, minimum width=31mm] (tok) at (35,0) {};
  \node[systok] (s0) at (22.5,0) {};
  \foreach \i/\x/\y in {1/27/6, 2/31.5/7.2, 3/36/7.8, 4/40.5/8.2, 5/27/0, 6/31.5/0, 7/36/0, 8/40.5/0, 9/27/-6, 10/31.5/-7.2, 11/36/-7.8}
    \node[bus] (b\i) at (\x,\y) {};
  \node[busflag] (b12) at (40.5,-8.2) {};
  \draw[line width=0.4pt, draw=black!55] (s0) -- (b1) -- (b2) -- (b3) -- (b4) (s0) -- (b5) -- (b6) -- (b7) -- (b8) (s0) -- (b9) -- (b10) -- (b11) -- (b12);
  \node[zonetok] (z1) at (46,7.6) {};
  \node[zonetok] (z2) at (46,0) {};
  \node[zonetok] (z3) at (46,-7.6) {};
  \draw[densely dotted, line width=0.4pt, draw=cB2] (b4) -- (z1) (b8) -- (z2) (b12) -- (z3);
  \node[lbl] at (35,-14.2) {residual tokens:\\$N$ bus, $Z$ zone, 1 system};
  \foreach \n in {mf, me, md, yy} \draw[msg] (\n.east) -- (\n.east -| tok.west);
  \node[blk, minimum width=19mm] (prgt) at (66,0) {};
  \node[layer] (l1) at (66,-7.8) {layer 1};
  \node[layer] (l2) at (66,-3.2) {layer 2};
  \node[layer] (l3) at (66,1.4) {layer $L$};
  \node[font=\scriptsize, align=center] at (66,7.4) {attention\\$+\,\beta_{\rho(i,j)}$};
  \node[lbl] at (66,-14.2) {PRGT with\\structural bias};
  \draw[msg] (tok.east) -- (prgt.west);
  \node[head] (h1) at (89.5,7.2) {claim scores $s(c)$};
  \node[head] (h2) at (89.5,0) {reconstructions};
  \node[head] (h3) at (89.5,-7.2) {root posterior};
  \node[lbl] at (89.5,-14.2) {three heads};
  \foreach \h in {h1, h2, h3} \draw[msg] (prgt.east |- \h) -- (\h.west);
  \node[guard, minimum width=18.5mm, minimum height=14mm, font=\scriptsize\bfseries] (cp) at (112.5,7.2) {Mondrian\\conformal\\p-values, BH};
  \draw[msg] (h1.east) -- (cp.west);
  \node[guard, minimum width=18.5mm, minimum height=14mm, font=\scriptsize\bfseries] (att) at (134.5,7.2) {attribute,\\ground,\\re-query};
  \draw[msg] (cp.east) -- (att.west);
  \node[lbl] at (123.5,15.6) {flags};
  \draw[msg] (h2.east) -| ([xshift=-3mm]att.south);
  \draw[msg] (h3.east) -| ([xshift=3mm]att.south);
  \node[guard, minimum width=18.5mm, minimum height=14mm, font=\scriptsize\bfseries] (rep) at (156.5,7.2) {repair QP\\with margin\\$m_\beta$};
  \draw[msg] (att.east) -- (rep.west);
  \node[lbl] at (145.5,15.6) {D still flagged};
  \node[plant, minimum width=10mm, minimum height=14mm, inner sep=1pt] (pl) at (174.8,7.2) {Feeder};
  \draw[msg] (rep.east) -- node[lbl, above] {$\bm{u}_t$} (pl.west);
  \draw[msg, draw=cB2] (cp.north) -- ++(0,7.4) -| node[lbl, pos=0.25, above, text=cB2] {no flag: execute the action of D} (pl.north);
  \draw[msg, draw=cOurs, densely dashed] ([xshift=6.5mm]att.south) -- ++(0,-22.5) -| node[lbl, pos=0.25, below, text=cOurs] {grounded claims back to E and D (one re-query)} (7,-17.5);
\end{tikzpicture}}
\caption{Architecture of GridAnchor.}
\label{fig:architecture}
\end{figure*}
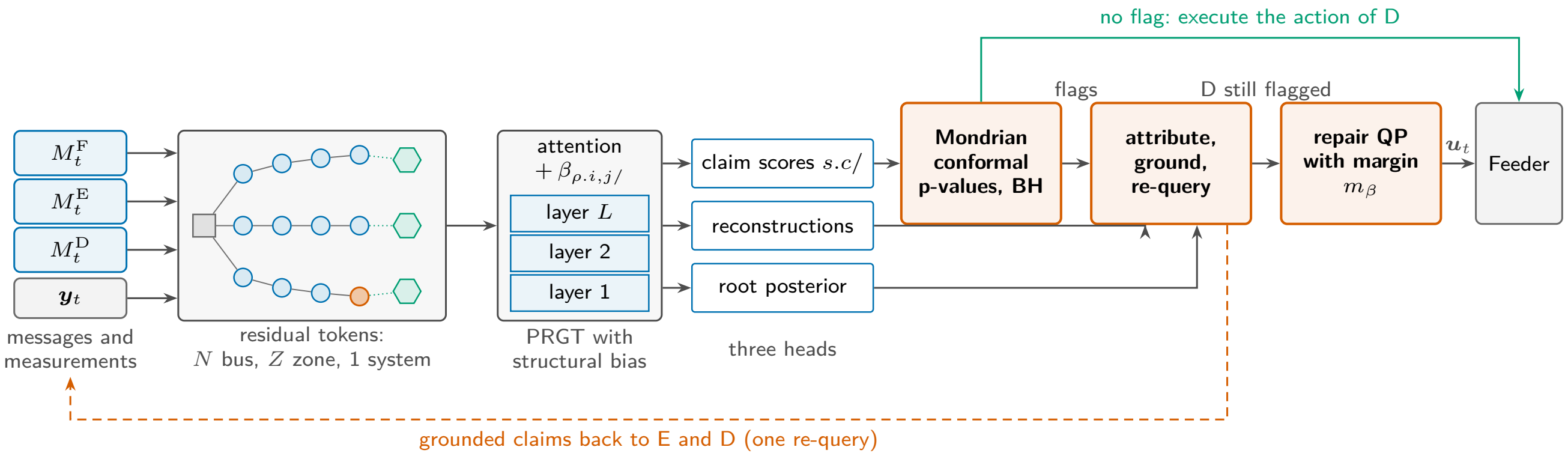

\Cref{fig:architecture} shows the verifier and its place between the agents and the plant.
At every step, GridAnchor converts the three messages and the measurements into physics-residual tokens on buses, zones and the system, and scores every claim with the PRGT.
It turns the scores into conformal p-values and applies the Benjamini-Hochberg (BH) procedure \cite{benjamini1995controlling}.
If no claim is flagged, the action of D is executed unchanged.
Otherwise the verifier attributes the root agent, grounds the flagged claims, queries the downstream agents again and, if the dispatch message is still flagged, repairs the action with a quadratic program (QP).

\subsection{Physics Model}
The verifier relies on a linearization of the power flow in the spirit of the DistFlow equations for radial feeders \cite{baran1989network}:
\begin{equation}
\bm{v} \approx \bm{v}^{\mathrm{ref}} + \bm{S}_p(\bm{p} - \bm{p}^{\mathrm{ref}}) + \bm{S}_q(\bm{q} - \bm{q}^{\mathrm{ref}}),
\label{eq:linear}
\end{equation}
where $\bm{S}_p, \bm{S}_q \in \mathbb{R}^{N \times N}$ are the sensitivities of the voltage magnitudes to the active and reactive injections.
They are computed once by central finite differences of the AC power flow around a mid-load reference point.
An action changes the injections by $\Delta\bm{p}(\bm{u}) $, obtained from \eqref{eq:injection}, so that the voltage after the action is predicted as $\hat{\bm{v}} + \bm{S}_p \Delta\bm{p}(\bm{u})$ from a voltage estimate $\hat{\bm{v}}$.
The electrical distance between two buses is $d_{ij} = S_{p,ii} + S_{p,jj} - 2S_{p,ij}$, and the zones are obtained by clustering the rows of $\bm{S}_p$.
Feeder losses are approximated by a quadratic function of the net zone injections fitted on training data.

\subsection{Physics-Residual Tokens}
\label{sec:tokens}
For each step the verifier computes residuals that vanish, up to noise and linearization error, when the claims agree with physics and with the measurements.
Let $\bm{p}^{\mathrm{F}}$ and $\bm{q}^{\mathrm{F}}$ be the injections implied by the claims of F through \eqref{eq:injection} with the previous action, and $\bm{v}^{\mathrm{F}}$ the corresponding voltages from \eqref{eq:linear}, shifted to the measured substation voltage.
With $\hat{\bm{v}}^{\mathrm{E}}$ the voltage claims of E and $v^{\mathrm{m}}_n$ the reading of meter $n \in \mathcal{M}$, the bus-level residuals are
\begin{align}
r^{\mathrm{volt}}_n &= \hat{v}^{\mathrm{E}}_n - v^{\mathrm{F}}_n, &
r^{\mathrm{meas}}_n &= \hat{v}^{\mathrm{E}}_n - v^{\mathrm{m}}_n, &
r^{\mathrm{link}}_n &= v^{\mathrm{m}}_n - v^{\mathrm{F}}_n,
\label{eq:busres}
\end{align}
where the last two exist only at metered buses.

The envelope residual $r^{\mathrm{env}}_n$ measures by how much a renewable claim exceeds the largest output seen in training for the same month and quarter-hour, or a load claim leaves its plausible range.
The temporal residual $r^{\mathrm{temp}}_n$ is the absolute change of a claim of F or E with respect to the previous grounded claim kept by the verifier.
The cross-sectional residual $r^{\mathrm{sub}}_n$ needs no memory.
The load, renewable and voltage claims of a step are each standardized and projected onto the principal subspace that explains 99.9\% of the variance of clean messages, and the residual is the part outside it.
It is close to zero for a coherent message, because bus loads follow a few shared profiles and bus voltages a few aggregate injection patterns.

At the system level, the balance residual compares the measured substation power $P_0$ with the claims of F:
\begin{equation}
r^{\mathrm{bal}} = P_0 - \Bigl(-\textstyle\sum_{n} p^{\mathrm{F}}_n + \hat{P}_{\mathrm{loss}}(\bm{p}^{\mathrm{F}}, \bm{q}^{\mathrm{F}})\Bigr).
\label{eq:bal}
\end{equation}
The rationale residuals are $r^{X} = X - \max_n \tilde{v}_n$ and $r^{C} = C - \sum_z \kappa_z \sum_{n \in z} \hat{p}^{\mathrm{R}}_n$, where $\tilde{\bm{v}} = \hat{\bm{v}}^{\mathrm{E}} + \bm{S}_p\Delta\bm{p}(\bm{u}_{\mathrm{D}})$ is the voltage predicted after the proposed action $\bm{u}_{\mathrm{D}}$.
At the zone level, $r^{\mathrm{soc}}_z$ measures violations of the power rating, of the curtailment range and of the bounds on the state of charge implied by $b_z$ and the measured state of charge.
The residual $r^{\mathrm{act}}_z$ is the largest excursion of $\tilde{\bm{v}}$ beyond the band among the buses of the zone.

Each residual is divided by a robust scale estimated on clean training data and compressed by a signed logarithm, which keeps gross errors such as unit slips bounded.
The scale is $1.4826$ times the median absolute deviation for the physical residuals.
The temporal and cross-sectional residuals, whose tails differ strongly from bus to bus, are divided by their 99.5\% clean quantile at the same bus.
One token per bus carries the two claims of F, the claim of E, the meter reading with its availability mask, the bus-level residuals and static attributes (capacities and the self-sensitivity $S_{p,nn}$).
One token per zone carries the action, the measured state of charge, the zone forecasts and the zone-level residuals, and one system token carries the substation measurements, $r^{\mathrm{bal}}$, $r^{X}$, $r^{C}$ and calendar encodings.
A step is therefore described by $N + Z + 1$ tokens.

\subsection{Physics-Residual Graph Transformer}
The PRGT is a pre-norm Transformer \cite{vaswani2017attention} over these tokens with $L$ layers, width $d$ and $H$ heads.
Like a graph neural network \cite{wu2021comprehensive}, it respects the feeder topology, through a bias on the attention logits:
\begin{equation}
a^{(h)}_{ij} = \frac{(\bm{W}^{(h)}_Q \bm{x}_i)^{\top}(\bm{W}^{(h)}_K \bm{x}_j)}{\sqrt{d/H}} + \beta^{(h)}_{\rho(i,j)},
\end{equation}
where $\bm{x}_i$ is the embedding of token $i$ and $\rho(i,j)$ is the relation between tokens $i$ and $j$.
For two bus tokens the relation is one of 16 quantile buckets of the electrical distance $d_{ij}$; other relations encode whether a bus belongs to a zone, zone pairs and pairs involving the system token.
The learned scalars $\beta^{(h)}_{\rho}$ are the only place where the topology enters, so the model has no parameter tied to a bus identity or to $N$.
Zone labels are permuted at random during training for the same reason.

Three heads read the final embeddings of the tokens.
The claim head outputs one logit $s(c)$ per claim: three per bus token (load, renewable and voltage), two per zone token (storage and curtailment) and two on the system token ($X$ and $C$).
The reconstruction head outputs a clean value for each claim of F and E as a convex combination of the claim, the previous grounded claim and, for voltages, the linear-model value, plus a bounded correction.
The root head outputs a posterior over $\{\text{none}, \mathrm{F}, \mathrm{E}, \mathrm{D}\}$ from the system token.

\subsection{Training by Taxonomy-Driven Synthesis}
Labeled hallucinations are rare in operating logs, so the PRGT is trained on synthetic episodes generated by CPH-Bench on the training days.
An episode corrupts the message of one agent according to one of the eight types, the downstream agents consume the corrupted message, and the plant simulator computes the labels of \cref{def:cph}, propagated hallucinations included.
The training loss combines four terms with fixed weights:
\begin{equation}
\mathcal{L} = \mathcal{L}_{\mathrm{focal}} + \lambda_{\mathrm{rec}}\mathcal{L}_{\mathrm{rec}} + \lambda_{\mathrm{phys}}\mathcal{L}_{\mathrm{phys}} + \lambda_{\mathrm{root}}\mathcal{L}_{\mathrm{root}}.
\label{eq:loss}
\end{equation}
In the order of \eqref{eq:loss}, the terms are a focal binary cross-entropy on the claim logits, a Huber loss between reconstructions and true values, a physics-consistency term \cite{karniadakis2021physics} and the cross-entropy of the root head.
The physics term is the norm of the residuals \eqref{eq:busres} and \eqref{eq:bal} recomputed with the reconstructed claims, which is differentiable because these residuals are linear in the claims.

Claims whose hallucination is natural, that is, caused by the agent itself on correct inputs, carry no weight in the focal term.
The synthesis cannot control them, and most leave no physical trace at the time of verification.
Training follows two curricula, one on severity and one on memory.
High-severity episodes are presented alone during the first four of the ten epochs, after which all severities are mixed.
During the last five epochs, half of each batch reads a memory built from the stochastic flags and reconstructions of the verifier itself.
The temporal residual thus learns on the partially corrected memory that it meets in operation.

\subsection{Conformal Calibration and Multiple Testing}
\label{sec:conformal}
The logits are not probabilities, and their distribution differs across claim types.
GridAnchor therefore converts them into split-conformal p-values \cite{vovk2022algorithmic,angelopoulos2023conformal} within six Mondrian classes $k$: F-load, F-renewable, E-voltage, D-storage, D-curtailment and D-rationale.
For each class, the calibration set $\{s_1, \dots, s_{n_k}\}$ contains the scores of the clean claims of the calibration days, and the p-value of a test claim $c$ of class $k$ is
\begin{equation}
p(c) = \frac{1 + \bigl|\{ i : s_i \ge s(c) \}\bigr|}{n_k + 1}.
\label{eq:pvalue}
\end{equation}

\begin{proposition}[Finite-sample false-alarm control]
\label{prop:far}
If, within a Mondrian class, the calibration scores and the score of a clean test claim are exchangeable, then $\Pr\{p(c) \le \alpha\} \le \alpha$ for every $\alpha \in (0,1)$.
If the scores are almost surely distinct, then also $\Pr\{p(c) \le \alpha\} \ge \alpha - 1/(n_k+1)$.
\end{proposition}

At each step the verifier tests all $m$ claims of the three messages together.
It sorts their p-values, $p_{(1)} \le \dots \le p_{(m)}$, finds the largest $i$ such that $p_{(i)} \le q\,i/m$ and flags the $i$ claims with the smallest p-values \cite{benjamini1995controlling}.
A message is flagged if at least one of its claims is flagged, and an unparsable message is always flagged.

\begin{proposition}[Per-step false discovery control]
\label{prop:fdr}
Assume that the clean scores are independent and identically distributed within each class and that the test claims are independent given the calibration data.
Then the conformal p-values \eqref{eq:pvalue} are positively regression dependent on the subset of clean claims within a class \cite{bates2023testing} and independent across classes, and the BH procedure at level $q$ yields a false discovery rate of at most $q\,m_0/m \le q$, where $m_0$ is the number of clean claims.
\end{proposition}

The claims of one step share the same physics, so the independence required by \cref{prop:fdr} is an idealization, and we treat the proposition as a design rationale.
\Cref{prop:far} needs only exchangeability within a class, which can fail under seasonal drift or after a change of an agent.
Both are therefore also checked empirically.

\subsection{Attribution and Grounded Re-Query}
\label{sec:mitigation}
When a claim is flagged, the root agent is the most probable agent under the root head among the agents with a flagged claim.
The topological rule used for comparison takes the first flagged agent in the order F, E, D.
Mitigation then follows the chain from F to D.
Flagged claims of F are replaced by the last value of the same claim that passed verification, and E is run again on the grounded forecast and verified again.
Claims of E that remain flagged are replaced by their reconstructions.
If any input of D has changed, D is queried once more with the grounded inputs and its new message is verified.

For the temporal residual of the next step, the verifier keeps a memory in which every flagged claim is replaced by its reconstruction.
A reconstruction stays close to a claim flagged by mistake, so that a false alarm cannot freeze the temporal reference.
The grounding rule for the downstream agents, last verified value for F and reconstruction for E, was the safest of three rules on the closed-loop validation days.

\subsection{Physics-Grounded Repair With a Conformal Margin}
\label{sec:repair}
If the message of D is still flagged, its action $\bm{u}_{\mathrm{D}}$ is projected onto a tightened linear model of the constraints:
\begin{equation}
\begin{aligned}
\min_{\bm{u},\,\bm{s}\ge \bm{0}}\quad & (\bm{u}-\bm{u}_{\mathrm{D}})^{\top}\bm{W}(\bm{u}-\bm{u}_{\mathrm{D}}) + \rho\,\bm{1}^{\top}\bm{s}\\
\text{s.t.}\quad & v_{\min} + m_\beta - \bm{s} \le \hat{\bm{v}} + \bm{S}_p\Delta\bm{p}(\bm{u}) \le v_{\max} - m_\beta + \bm{s},\\
& \bm{u} \in \mathcal{U}(\bm{e}_t).
\end{aligned}
\label{eq:repair}
\end{equation}
Here $\hat{\bm{v}}$ is the grounded voltage estimate, and $\mathcal{U}(\bm{e}_t)$ collects the power ratings, the curtailment range and the bounds on the state of charge.
The matrix $\bm{W}$ is diagonal and positive definite (inverse squared ratings for storage, one for curtailment fractions), and $\rho = 10^4$ penalizes the slack $\bm{s}$.
Filtering a learned action through a constrained program follows predictive safety filters \cite{wabersich2021predictive}, here with verified inputs and a calibrated tightening.

The margin $m_\beta$ is the conformal $(1-\beta)$ quantile of the end-to-end error
\begin{equation}
\delta(\bm{u}) = \max_n \bigl|v_n(\bm{u}) - \hat{v}_n - (\bm{S}_p\Delta\bm{p}(\bm{u}))_n\bigr|
\label{eq:delta}
\end{equation}
between the AC voltages $\bm{v}(\bm{u})$ and the prediction, which covers estimation, forecast and linearization errors at once.
The errors are measured where the repair acts: closed-loop runs on the calibration days, with injected episodes and the full mitigation sequence, record $\delta(\bm{u}^\star)$ at every repaired step.
They thus come from the same selection by the detector, the same grounding of $\hat{\bm{v}}$ and the same program as at run time, and each grounding rule receives its own margin.
These runs start from an offline margin computed at solutions of \eqref{eq:repair} for wrong proposals on the calibration log (random perturbations of the expert action and the altered dispatch messages of an injected chain).

\begin{proposition}[Feasibility and minimality of the repair]
\label{prop:repair}
Assume that the error $\delta(\bm{u}^\star)$ at the repaired action $\bm{u}^\star$ is exchangeable with the calibration errors used to compute $m_\beta$, and that \eqref{eq:repair} is solved with $\bm{s} = \bm{0}$.
Then the AC voltages under the repaired action lie in $[v_{\min}, v_{\max}]$ with probability at least $1-\beta$.
In addition, the repaired action is the unique feasible action that minimizes the $\bm{W}$-norm of the modification of $\bm{u}_{\mathrm{D}}$.
\end{proposition}

Because the calibration errors arise from the same selection, grounding and optimization as $\delta(\bm{u}^\star)$, exchangeability reduces to the requirement that the calibration days resemble the test days.
The coverage $\Pr\{\delta(\bm{u}^\star) \le m_\beta\}$ is checked at the repaired steps of the test runs.
The repair, finally, cannot fix a voltage estimate that is wrong by more than the margin, which is why the grounding rule matters.

\subsection{Algorithm and Complexity}
\Cref{alg:gridanchor} summarizes one verification step of the layer in pseudocode.
The residuals cost two products with the $N \times N$ sensitivity matrices, the PRGT $O(L\,(N+Z+1)^2 d)$ operations and the p-values one binary search per claim.
The QP has $3Z$ decision variables and $N$ slack variables.

\begin{algorithm}[t]
\caption{GridAnchor, one step $t$}
\label{alg:gridanchor}
\begin{algorithmic}[1]
\Require messages $M^{\mathrm{F}}_t, M^{\mathrm{E}}_t, M^{\mathrm{D}}_t$; measurements $\bm{y}_t$; previous action $\bm{u}_{t-1}$; verifier memories; calibration scores per class; level $q$; margin $m_\beta$
\State compute the residual tokens
\State run the PRGT: claim scores, reconstructions, root posterior
\State compute the conformal p-values \eqref{eq:pvalue}
\State apply the BH procedure at level $q$ to the claims of the step
\If{no claim is flagged}
  \State $\bm{u}_t \gets \bm{u}_{\mathrm{D}}$
\Else
  \State attribute the root agent
  \State replace the flagged claims of F by their last verified values
  \State run E again on grounded inputs and verify its message
  \State replace the claims of E that remain flagged by their reconstructions
  \State query D again on grounded inputs and verify its message
  \If{the message of D is flagged}
    \State $\bm{u}_t \gets$ solution of the repair QP \eqref{eq:repair}
  \Else
    \State $\bm{u}_t \gets$ action of the new message of D
  \EndIf
\EndIf
\State update both memories, log the decision, execute $\bm{u}_t$
\end{algorithmic}
\end{algorithm}

\section{Experimental Setup}
\label{sec:setup}

\subsection{Testbed}
The main feeder is the SimBench grid 1-MV-rural with its one-year profiles at a resolution of 15~min \cite{meinecke2020simbench}; \cref{tab:testbed} summarizes it together with the agents, the data and the measured training times.
It has \resBusMV{} buses at 20~kV behind two transformers, 96 loads with a peak of \resPeakLoadMW~MW, \resResCapMW~MW of curtailable renewable generation and \resStorageMWh~MWh of storage rated at \resStorageMW~MW.
The storage units of each of the $Z = \resZones$ zones act as one aggregated unit, and the zones, a k-means clustering of the normalized rows of $\bm{S}_p$, follow the feeder branches.

Without control, \resUncontrolledStepPct\% of the steps of the year have at least one bus outside the band $[0.95, 1.05]$~p.u., with voltages up to \resUncontrolledVmax~p.u., mainly because of the wind farms connected at medium voltage.
These are therefore curtailable, together with the aggregated low-voltage generation and the photovoltaic plants, as in feed-in management, whereas biomass and hydro units (\resNcCapMW~MW) are only telemetered.
The plant is simulated with a vectorized fixed-point AC power flow on the admittance matrix built by pandapower \cite{thurner2018pandapower}, whose Newton-Raphson solution it matches within $\resPfMaxDev$~p.u.

\begin{table}[t]
\centering
\caption{Testbed, agents and data.}
\label{tab:testbed}
\setlength{\tabcolsep}{3pt}
\footnotesize
\begin{tabular}{@{}p{16mm}p{69mm}@{}}
\toprule
\multicolumn{2}{@{}l}{\textit{Feeders (SimBench)}} \\
Main & 1-MV-rural: \resBusTotal{} buses (\resBusMV{} at 20~kV), \resResCapMW~MW curtailable renewables, \resStorageMWh~MWh storage, \resUncontrolledStepPct\% of steps with a voltage violation without control \\
Transfer & 1-MV-semiurb: \resTBusTotal{} buses (\resTBusMV{} at 20~kV), \resTResCapMW~MW curtailable renewables, \resTStorageMWh~MWh storage, \resTUncontrolledStepPct\% of steps with a violation without control \\
\midrule
\multicolumn{2}{@{}l}{\textit{Agents and verifier (main feeder)}} \\
F & shared MLP, two hidden layers of 128 units, bus embedding; \resFParamsK{}k parameters; trained in \resFTrainS~s; validation RMSE \resFRmsePct\% of the rating \\
E & MLP, three hidden layers of 256 units; \resEParamsK{}k parameters; trained in \resETrainS~s; validation RMSE \resERmseMpu$\times 10^{-3}$~p.u. \\
D & decoder-only Transformer, 4 layers, width 128, context 256; \resDParamsM{}M parameters; \resDTrainMin~min; \resDParsePct\% parsable outputs, storage error \resDBMaePct\% of rating, curtailment error \resDKMaePp~points \\
PRGT & 3 layers, width 64, 4 heads; \resPrgtParamsK{}k parameters; \resPrgtTrainMin~min per seed \\
\midrule
\multicolumn{2}{@{}l}{\textit{Data and measurements}} \\
Splits & 13 blocks of 28 days: \resTrainDays{} training, \resCalDays{} calibration, \resValDays{} validation and \resTestDays{} test days, with gap days; \resClosedDays{} closed-loop test days \\
Sensors & substation $P$, $Q$ (0.5\%) and $V$ (0.1\%); \resMeters{} voltage meters (0.2\%, 2\% missing); state of charge (0.5 points); telemetry of non-controllable units (1\%) \\
\bottomrule
\end{tabular}

\end{table}

The year is cut into 13 blocks of 28 days, with gap days against leakage.
In each block, days 1 to 16 serve for training, days 18 to 20 for conformal calibration, days 21 to 23 for validation and days 25 to 28 for testing.
Open-loop experiments use the \resTestDays{} test days.
Closed-loop runs use the \resClosedCalDays{} calibration days, the \resClosedValDays{} validation days and the \resClosedDays{} test days of blocks 1, 4, 7 and 10, one block per season.
Voltage meters cover \resMeters{} buses: the farthest bus of each zone, the buses of the generators connected at medium voltage and a farthest-point completion.

\subsection{Agents}
An expert controller provides the reference trajectory and the labels of the imitation data.
At each step it solves a QP over $\bm{u}_t$ with the true state and the linear model \eqref{eq:linear}.
It minimizes the cost of imported energy, a curtailment compensation, a storage throughput cost and the deviation from a daily state-of-charge reference.
Its constraints are the device limits and the voltage band tightened by 0.004~p.u.; it needs voltage slack on \resExpertSlackPct\% of the training steps and curtails \resExpertCurtPct\% of the available renewable energy.

The agents F and E are multilayer perceptrons, and D is a small language model (SLM) trained from scratch, in line with the argument that small specialized models fit repetitive agentic tasks \cite{belcak2025small}.
Its prompt lists, for each zone, the measured state of charge, the zone forecasts of F, the maximum zone voltage claimed by E and the previous setpoints.
Its completion gives $b_z$ and $\kappa_z$ for each zone, then $X$ and $C$, under a strict grammar.
The SLM imitates the expert on \resDNTrain{} sequences for \resDEpochs{} epochs, two thirds of them from states with perturbed state of charge and setpoints, and decodes greedily.

\subsection{CPH-Bench Injection Protocol}
In open loop the plant follows the expert trajectory, and the three agents produce their messages at every step from the corresponding state.
An idle step starts an episode with probability 0.10 on test days and 0.40 on training days, so that about half of the training steps contain an active episode.
The episode targets F, E or D with probabilities 0.4, 0.3 and 0.3 and draws one of the types applicable to that agent uniformly, a severity level and a persistence from 1 to 4 steps.
The corrupted message is consumed by the downstream agents, so cascades arise without being scripted.
Each claim receives one of four labels from \cref{def:cph}: injected, propagated (by a counterfactual run of the agent with ground-truth inputs), natural or clean.

In the detection experiments, a message is positive if it contains an injected or propagated hallucination and negative if it contains no hallucination, so that a message altered by less than the tolerance is negative.
Messages whose only hallucinations are natural are set aside and studied separately, because the protocol does not control them and most natural forecast errors leave no trace in the measurements of the same step.
Counting them as positives gives secondary scores.

\subsection{Baselines}
Seven baselines share the code base of GridAnchor.
B1 applies plausibility rules (static ranges, the renewable envelope and ramp limits).
B2 is a linearized weighted-least-squares estimator of the bus injections with the claims of F and E as pseudo-measurements, scored by normalized residuals and by the chi-square statistic \cite{abur2004power}.
B3 is an Isolation Forest \cite{liu2008isolation} and B4 a dense autoencoder, both fitted per agent on message vectors with measurement context, as representatives of the data-driven detectors surveyed in \cite{musleh2020survey}.
B5 is a self-consistency check: deep ensembles of five forecasters and five estimators \cite{lakshminarayanan2016simple} and five stochastic samples of the SLM at temperature 0.7, in the manner of sampling-based hallucination detectors \cite{manakul2023selfcheckgpt,farquhar2024detecting}.
B6 is a digital-twin check in the spirit of external verifiers \cite{kambhampati2024llms}: an AC power flow with the injections claimed by F and the action of D flags E and D when they contradict the simulation.
B7 is a predictive safety filter \cite{wabersich2021predictive} that projects every action with \eqref{eq:repair} without any detection.

GridAnchor learns from labels defined by the tolerances $\epsilon_k$ of \cref{def:cph}, which the other detectors do not know.
Two variants therefore give them to B4 and B5, the strongest data-driven baselines.
A claim is then scored only by the part of its deviation that exceeds $d\,\epsilon_k$, measured from the reconstruction of B4 or from the center of the samples of B5.
All detectors share the same Mondrian conformal calibration and BH step, so that differences reflect the quality of the scores, and each was tuned on validation data within ten configurations, the deadband factor $d$ included.
In closed loop, the detectors B1 to B6 use the mitigation sequence and the repair margin of GridAnchor; having no reconstruction head, they replace every flagged claim by its last verified value.

\subsection{Metrics and Statistics}
The primary detection metric is the area under the precision-recall curve (AUPRC) over messages.
Each detector yields one statistic per message, the largest score among its claims or, for B2 to B4, its own message score when that validated better.
The statistic is converted into a conformal p-value per agent, so that the messages of the three agents are ranked together.
Scores beyond the largest calibration score, which a p-value cannot resolve, are ranked by their relative exceedance of that maximum, for the ranking metrics only.

We also report the area under the receiver operating characteristic curve (AUROC) and the true-positive rate (TPR) at a false-alarm rate (FAR) of 1\%.
For the decisions, we report the false discovery rate (FDR) of the BH step and the empirical FAR of the per-claim test.
The FAR is measured on the clean claims of the steps without an active episode, where the assumption of \cref{prop:far} can hold.
Closed-loop metrics are the share of bus-steps outside the voltage band, the worst overvoltage, the curtailed energy and the operating cost.
They also count the zone-steps beyond the state-of-charge bounds and the executed hallucinated actions, and give the share of episodes whose cascade does not reach the plant.

Results are means and standard deviations over \resSeeds{} seeds, which change the initialization of the PRGT, the injection sampling and the randomness of the baselines.
The seeds share the test days, so the day is the unit of every statistical comparison, and the daily value of a metric is averaged over the seeds.
Confidence intervals are 95\% percentile bootstrap intervals over days, and paired Wilcoxon signed-rank tests over days are Holm-corrected.

\subsection{Implementation}
The PRGT has \resPrgtParamsK{}k parameters and is trained with AdamW for \resPrgtEpochsMax{} epochs of \resPrgtEpochSteps{} sampled steps; the epoch with the best validation AUPRC after the severity curriculum is kept.
The levels are $q = 0.1$ and $\beta = 0.05$.
The offline margin is $\resMarginMpu \times 10^{-3}$~p.u. from \resMarginN{} errors at program solutions, and the margin of the selected grounding rule is $\resMarginClMpu \times 10^{-3}$~p.u. from \resMarginClN{} repaired steps of the calibration runs.
All experiments ran on two CPU cores without a graphics processor, with Python 3.13, PyTorch 2.15, pandapower 3.5, SimBench 1.6 and CVXPY 1.9.
One seed is used for the retrained ablations, the leave-one-type-out models and the in-domain transfer model, all trained with the full schedule, and the SLM of the transfer feeder is trained on \resTDNTrain{} sequences.

\section{Results}
\label{sec:results}

\subsection{Natural Hallucinations of the Agents}
\label{sec:census}
\Cref{tab:census} reports the hallucinations that the agents produce without any injection, at half, nominal and double tolerance, and the share that GridAnchor flags at $\alpha = 0.05$.
On the \resTestDays{} test days, the forecaster exceeds the tolerance on \resNatFLoadPct\% of its load claims, in \resNatFMsgPct\% of its messages, while the estimator stays within the tolerance on every voltage claim.
The SLM dispatcher emits a hallucinated message in \resNatDMsgPct\% of the steps, most often a stated total curtailment that its own action does not produce (\resNatDCPct\% of the steps).
Of these messages, \resNatDPropSharePct\% are propagated from a forecast error.

The verifier flags \resNatRecallDCOursPct\% of the \resNatCountDC{} unfaithful curtailment statements but only \resNatRecallFLoadOursPct\% of the natural load-forecast errors.
A forecast error on one bus leaves almost no trace in the measurements of the same step, so a detector that reads concurrent measurements cannot see it.
Self-consistency does not see it either, because ensemble members trained on the same data share the errors of the model they copy: B5 flags \resNatRecallFLoadBfivePct\% of these errors.

\begin{table}[t]
\centering
\caption{Natural hallucinations on test days.}
\label{tab:census}
\footnotesize
\setlength{\tabcolsep}{4pt}
\begin{tabular}{@{}lcccc@{}}
\toprule
 & \multicolumn{3}{c}{Rate (\%), tolerance scale} & Flagged \\
\cmidrule(lr){2-4}
Agent and form & $\times 0.5$ & $\times 1$ & $\times 2$ & (\%) \\
\midrule
F, load value & 3.315 & 0.463 & 0.016 & 26.9\std{5.4} \\
F, renewable value & 0.090 & 0.003 & 0.000 & 85.7\std{11.7} \\
E, voltage value & 0.000 & 0.000 & 0.000 & none \\
D, device limit & 0.000 & 0.000 & 0.000 & none \\
D, unsafe action & 0.020 & 0.020 & 0.000 & 100.0\std{0.0} \\
D, unfaithful $X$ & 1.242 & 0.000 & 0.000 & none \\
D, unfaithful $C$ & 1.322 & 0.361 & 0.000 & 100.0\std{0.0} \\
D, format & 0.000 & 0.000 & 0.000 & none \\
\bottomrule
\end{tabular}

\end{table}

\begin{table*}[t]
\centering
\caption{Detection of injected and propagated hallucinations on test days.}
\label{tab:detection}
\footnotesize
\setlength{\tabcolsep}{3.6pt}
\resizebox{\textwidth}{!}{
\begin{tabular}{@{}lccccccccc@{}}
\toprule
 & & \multicolumn{4}{c}{AUPRC $\uparrow$} & TPR at & FAR at & FDR at & Latency \\
\cmidrule(lr){3-6}
Detector & AUROC $\uparrow$ & all & F & E & D & 1\% FAR $\uparrow$ & $\alpha{=}0.05$ & $q{=}0.1$ & (ms) $\downarrow$ \\
\midrule
B1 Rules & 0.883\std{0.006} & 0.742\std{0.019} & 0.673\std{0.034} & 0.839\std{0.004} & 0.739\std{0.097} & 0.717\std{0.007} & 0.054\std{0.001} & 0.146\std{0.002} & \best{0.9} \\
B2 WLS-BDD & 0.924\std{0.004} & 0.576\std{0.038} & 0.357\std{0.061} & 0.874\std{0.071} & 0.605\std{0.052} & 0.576\std{0.008} & 0.051\std{0.001} & 0.110\std{0.007} & 1.5 \\
B3 IsoForest & 0.866\std{0.000} & 0.547\std{0.038} & 0.514\std{0.043} & 0.685\std{0.052} & 0.436\std{0.113} & 0.494\std{0.078} & 0.050\std{0.000} & 0.073\std{0.002} & 33.5 \\
B4 Autoencoder & 0.976\std{0.005} & 0.837\std{0.014} & 0.889\std{0.012} & 0.940\std{0.005} & 0.672\std{0.015} & 0.774\std{0.043} & 0.051\std{0.001} & 0.206\std{0.010} & 1.7 \\
B5 Self-consist. & \second{0.979}\std{0.001} & 0.830\std{0.033} & \best{0.986}\std{0.001} & 0.912\std{0.026} & 0.689\std{0.082} & 0.678\std{0.028} & 0.052\std{0.000} & 0.199\std{0.004} & 99.7 \\
B6 Digital twin & 0.645\std{0.034} & 0.458\std{0.076} & 0.051\std{0.001} & 0.870\std{0.066} & 0.662\std{0.081} & 0.561\std{0.041} & 0.018\std{0.000} & 0.055\std{0.001} & \second{1.1} \\
\midrule
B4 + tolerance & 0.973\std{0.004} & 0.811\std{0.056} & 0.891\std{0.020} & \second{0.944}\std{0.007} & \second{0.753}\std{0.014} & 0.813\std{0.029} & 0.007\std{0.001} & 0.163\std{0.007} & 1.7 \\
B5 + tolerance & 0.942\std{0.008} & \best{0.874}\std{0.020} & \second{0.956}\std{0.002} & 0.826\std{0.041} & 0.705\std{0.080} & \best{0.862}\std{0.036} & 0.000\std{0.000} & 0.033\std{0.004} & 84.9 \\
\midrule
GridAnchor (ours) & \best{0.984}\std{0.002} & \second{0.846}\std{0.033} & 0.856\std{0.033} & \best{0.967}\std{0.011} & \best{0.905}\std{0.026} & \second{0.814}\std{0.020} & 0.051\std{0.001} & 0.181\std{0.006} & 4.0 \\
\bottomrule
\end{tabular}
}
\end{table*}

\subsection{Detection of Injected and Propagated Hallucinations}
\label{sec:detection}
\Cref{tab:detection} compares the detectors on the test days (mean and standard deviation over \resSeeds{} seeds; best value in bold, second best underlined), where \resMsgPrevalencePct\% of the scored messages are positive.
GridAnchor reaches an AUPRC of \resAuprcOurs{} (95\% confidence interval \resAuprcOursCiLo{} to \resAuprcOursCiHi) and a TPR of \resTprOursPct\% at a FAR of 1\%.
The strongest baseline, \resBestBaselineName, reaches \resAuprcBestBaseline{} (\resAuprcBestCiLo{} to \resAuprcBestCiHi) and \resTprBestBaselinePct\%.
The confidence interval of the difference in AUPRC, \resAuprcDiffCiLo{} to \resAuprcDiffCiHi, contains zero, so the test days do not separate the two detectors.
Paired Wilcoxon signed-rank tests on the daily AUPRC over the \resWilcoxonPairs{} test days, with Holm correction over all comparisons, compare GridAnchor with each baseline.
GridAnchor is significantly better than \resWilcoxonNSigBase{} of the six baselines (corrected $p < 0.05$), but not better than B4 or B5.

B5 recomputes the forecast with five independently trained models, so an injected deviation stands out whatever its type, and it leads on F with \resAuprcFBfive, ahead of B4 (\resAuprcFBfour) and GridAnchor (\resAuprcFOurs).
GridAnchor leads on E with \resAuprcEOurs{} and on D with \resAuprcDOurs, where the best baseline reaches \resAuprcDBest{} (\resAuprcDBestName).
B4 and B5 have no physical model of an action, and resampling the SLM reproduces a message propagated from a hallucinated input.
The lowest per-agent AUPRC of GridAnchor, \resAuprcMinAgentOurs, exceeds that of every baseline (at most \resAuprcMinAgentBest), with one model and no copy of the agents.

\begin{figure}[t]
\centering
\begin{tikzpicture}
\begin{axis}[paperstyle, height=0.7\columnwidth, ybar=0.6pt, bar width=3.7pt, ymin=0, ymax=1.08, xmin=0.4, xmax=8.6,
  xtick={1,...,8}, xticklabels={H1,H2,H3,H4,H5,H6,H7,H8}, ytick={0,0.2,0.4,0.6,0.8,1.0},
  ylabel={Message-level AUPRC}, xlabel={Hallucination type}, xmajorgrids=false,
  unbounded coords=discard, every axis plot/.append style={line width=0.3pt},
  legend columns=2, legend style={at={(0.5,1.03)}, anchor=south, legend cell align=left, /tikz/every even column/.append style={column sep=3pt}},
  legend image code/.code={\draw[#1] (0cm,-0.09cm) rectangle (0.2cm,0.12cm);},
  error bars/y dir=both, error bars/y explicit, error bars/error bar style={line width=0.35pt, black!75},
  error bars/error mark options={rotate=90, mark size=1pt, line width=0.35pt, black!75}]
\addplot[fill=cOurs, draw=cOurs!80!black] table[x=idx, y=GridAnchor, y error=GridAnchorsd, col sep=comma] {figures/data/fig3_pertype.csv};
\addplot[fill=\resTopAColor!75, draw=\resTopAColor!80!black] table[x=idx, y=\resTopAKey, y error=\resTopAKey sd, col sep=comma] {figures/data/fig3_pertype.csv};
\addplot[fill=\resTopBColor!75, draw=\resTopBColor!80!black] table[x=idx, y=\resTopBKey, y error=\resTopBKey sd, col sep=comma] {figures/data/fig3_pertype.csv};
\addplot[fill=\resTopCColor!75, draw=\resTopCColor!80!black] table[x=idx, y=\resTopCKey, y error=\resTopCKey sd, col sep=comma] {figures/data/fig3_pertype.csv};
\legend{GridAnchor, \resTopAName, \resTopBName, \resTopCName}
\end{axis}
\end{tikzpicture}
\caption{Message-level AUPRC per hallucination type.}
\label{fig:pertype}
\end{figure}
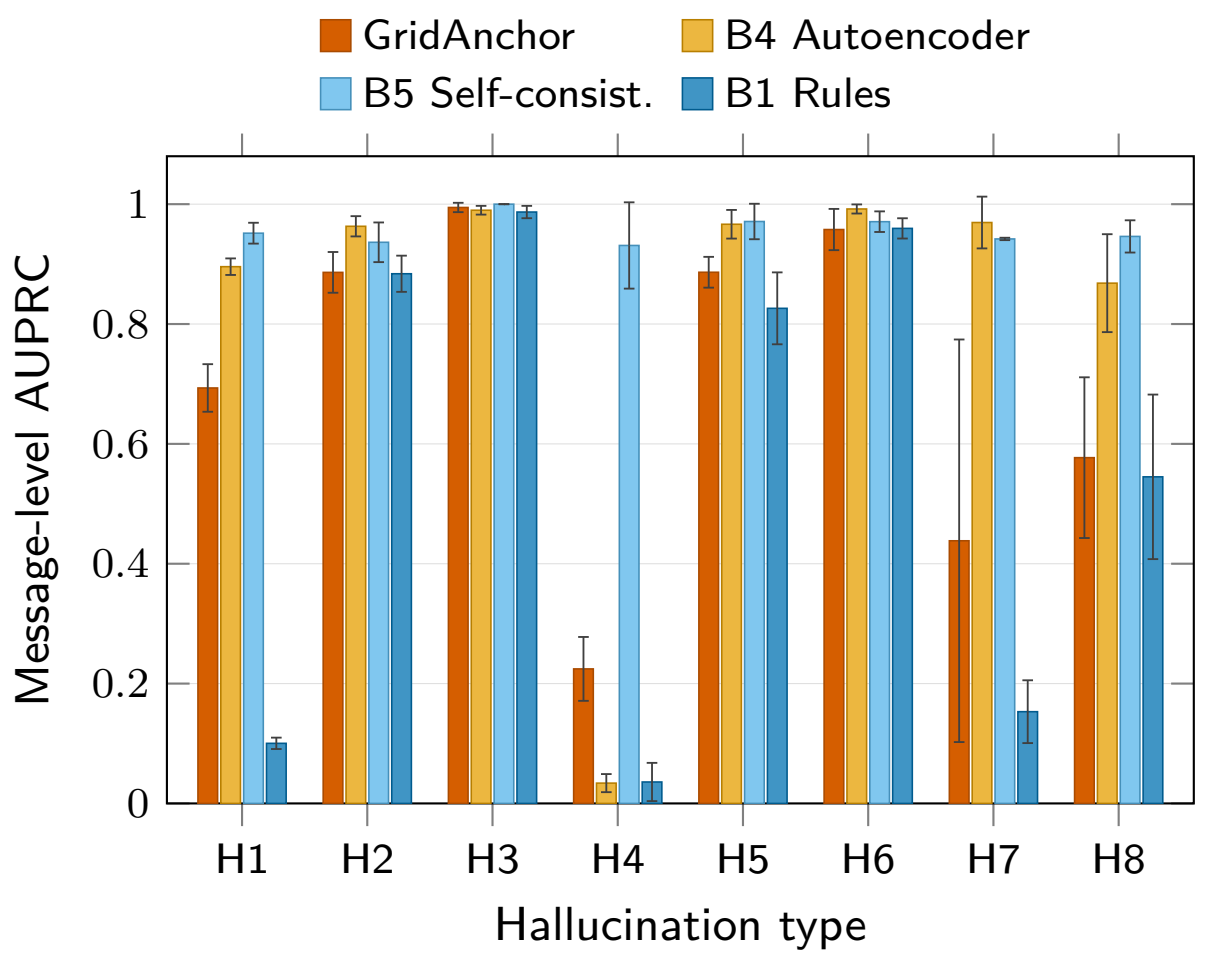

\Cref{fig:pertype} breaks the comparison down by type for GridAnchor and the three strongest baselines (mean and standard deviation over seeds), with the positives of one type scored against the messages of steps without any episode.
In this view GridAnchor is behind the best baseline on every type.
B5 stays above \resTypeMinBfive{} on all of them, because a recomputed message shares neither the delay of a stale replay nor the offset of a drift.
The weakest types of GridAnchor are stale replays (H4, \resTypeHFourOurs), drifts (H7, \resTypeHSevenOurs) and unfaithful rationales (H8, \resTypeHEightOurs).

The two views differ because of the negatives that the per-type view leaves out.
At a FAR of 1\%, \resFpAlteredBfivePct\% of the false positives of B5 and \resFpAlteredBfourPct\% of those of B4 are messages that an episode altered by less than the tolerance, against \resFpAlteredPct\% for GridAnchor.
Such messages make up only \resNegAlteredPct\% of the negatives.
B4 and B5 detect that a message has changed, whereas GridAnchor, trained on the labels of \cref{def:cph}, more often separates a change within the tolerance from a hallucination.

The two variants that know the tolerances test this explanation.
With a deadband of $d = \resDeadbandBfiveTol$ tolerances, selected on validation days, B5 reaches an AUPRC of \resAuprcBfiveTol{} (\resAuprcBfiveTolCiLo{} to \resAuprcBfiveTolCiHi), above GridAnchor, and a TPR of \resTprBfiveTolPct\% at a FAR of 1\%.
The interval of the difference, \resAuprcDiffBfiveTolCiLo{} to \resAuprcDiffBfiveTolCiHi, still contains zero (corrected $p = \resWilcoxonBfiveTolP$).
GridAnchor keeps its lead on E and D, where this variant reaches \resAuprcEBfiveTol{} and \resAuprcDBfiveTol, and the deadband lowers the AUROC of B5 to \resAurocBfiveTol{} because it sets most clean scores to zero.
The autoencoder does not gain from a deadband (\resAuprcBfourTol{} with $d = \resDeadbandBfourTol$).
The lead of GridAnchor over the six baselines thus owes much to its knowledge of the tolerances, and its own contribution lies in the estimates and the actions, which physics constrains.
When natural hallucinations count as positives, the prevalence rises to \resMsgPrevalenceAllPct\% and every detector drops, \resAuprcAllBestName{} to \resAuprcAllBest{} and GridAnchor to \resAuprcAllOurs.

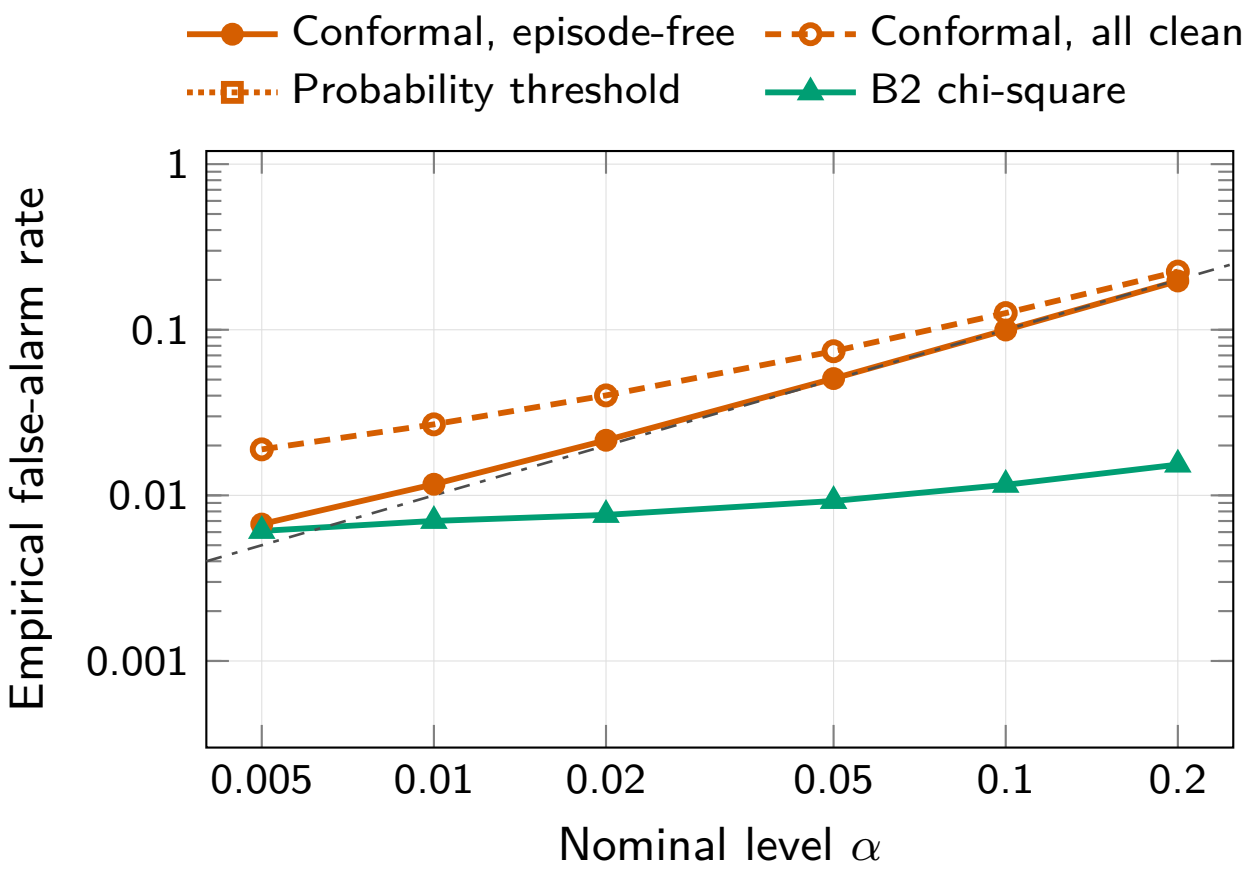
\begin{figure}[t]
\centering
\begin{tikzpicture}
\begin{axis}[paperstyle, height=0.66\columnwidth, xmode=log, ymode=log, xmin=0.004, xmax=0.25, ymin=0.0003, ymax=1.2,
  xtick={0.005,0.01,0.02,0.05,0.1,0.2}, xticklabels={0.005,0.01,0.02,0.05,0.1,0.2},
  ytick={0.001,0.01,0.1,1}, yticklabels={0.001,0.01,0.1,1},
  xlabel={Nominal level $\alpha$}, ylabel={Empirical false-alarm rate},
  legend columns=2, legend style={at={(0.5,1.03)}, anchor=south, legend cell align=left, /tikz/every even column/.append style={column sep=3pt}}]
\addplot[cOurs, mark=*, mark size=1.6pt] table[x=alpha, y=conformal, col sep=comma] {figures/data/fig4_calibration.csv};
\addplot[cOurs, densely dashed, mark=o, mark size=1.7pt, mark options={solid}] table[x=alpha, y=conformalall, col sep=comma] {figures/data/fig4_calibration.csv};
\addplot[cOurs, densely dotted, mark=square, mark size=1.6pt, mark options={solid}] table[x=alpha, y=fixed, col sep=comma] {figures/data/fig4_calibration.csv};
\addplot[cB2, mark=triangle*, mark size=1.9pt] table[x=alpha, y=chi, col sep=comma] {figures/data/fig4_calibration.csv};
\addplot[black!70, dashdotted, no marks, domain=0.004:0.25, samples=2, line width=0.5pt, forget plot] {x};
\legend{Conformal{,} episode-free, Conformal{,} all clean, Probability threshold, B2 chi-square}
\end{axis}
\end{tikzpicture}
\caption{Empirical versus nominal false-alarm rate.}
\label{fig:calibration}
\end{figure}

\subsection{Calibration}
\label{sec:calibration}
\Cref{fig:calibration} tests the false-alarm guarantee of \cref{prop:far} on test days for the conformal p-values of GridAnchor, a threshold on its output probability and the chi-square threshold of B2 (dash-dotted line: identity).
On the steps without an active episode, the empirical FAR of the conformal test is \resFarOursOnePct\%, \resFarOursFivePct\% and \resFarOursTenPct\% at nominal levels of 1\%, 5\% and 10\%, at most \resFarMaxGapPp{} percentage points from the target.
At a nominal 5\%, the probability threshold gives \resFarFixedFivePct\% and the chi-square threshold, which assumes Gaussian noise, \resFarChiFivePct\%.
Over all clean claims the conformal FAR is \resFarAllOursFivePct\%, because it reaches \resFarEpiOursFivePct\% on the clean claims that share a step with a hallucination.
Their residuals are computed with the hallucinated claims, so they are not exchangeable with the calibration claims.

The FDR of the BH decisions at $q = 0.1$ shows the same limit: \resFdrFreeOurs{} on the steps without an episode, \resFdrEpiOurs{} on the steps with one and \resFdrOurs{} in total.
On the steps with at least one rejection, the share of false claims among the flagged ones is \resFdrRejectingOurs{} on average.
The flags thus locate a hallucinated claim far less precisely than they detect a hallucinated message, and $q$ bounds no share of false flags during an episode.

\begin{figure}[t]
\centering
\begin{tikzpicture}
\begin{axis}[paperstyle, xmin=0.3, xmax=8.2, ymin=0, ymax=1.05, xtick={0.5,1,2,3,4,6,8}, xticklabels={0.5,1,2,3,4,6,8},
  xlabel={Drift magnitude $s$ (multiples of the clean error spread)}, ylabel={TPR at 1\% false-alarm rate},
  height=0.68\columnwidth, legend columns=2, legend style={at={(0.5,1.03)}, anchor=south, legend cell align=left, /tikz/every even column/.append style={column sep=3pt}}]
\newcommand{\band}[2]{%
  \addplot[draw=none, name path=lo#1, forget plot] table[x=s, y=#1lo, col sep=comma] {figures/data/fig5_severity.csv};
  \addplot[draw=none, name path=hi#1, forget plot] table[x=s, y=#1hi, col sep=comma] {figures/data/fig5_severity.csv};
  \addplot[fill=#2, fill opacity=0.16, draw=none, forget plot] fill between[of=lo#1 and hi#1];}
\band{GridAnchor}{cOurs}
\expandafter\band\expandafter{\resTopAKey}{\resTopAColor}
\expandafter\band\expandafter{\resTopBKey}{\resTopBColor}
\expandafter\band\expandafter{\resTopCKey}{\resTopCColor}
\addplot[cOurs, mark=*, mark size=1.2pt] table[x=s, y=GridAnchor, col sep=comma] {figures/data/fig5_severity.csv};
\addplot[\resTopAColor, mark=square*, mark size=1.1pt] table[x=s, y=\resTopAKey, col sep=comma] {figures/data/fig5_severity.csv};
\addplot[\resTopBColor, mark=triangle*, mark size=1.4pt] table[x=s, y=\resTopBKey, col sep=comma] {figures/data/fig5_severity.csv};
\addplot[\resTopCColor, mark=diamond*, mark size=1.4pt] table[x=s, y=\resTopCKey, col sep=comma] {figures/data/fig5_severity.csv};
\legend{GridAnchor, \resTopAName, \resTopBName, \resTopCName}
\end{axis}
\end{tikzpicture}
\caption{Detection of drifts versus drift magnitude.}
\label{fig:severity}
\end{figure}
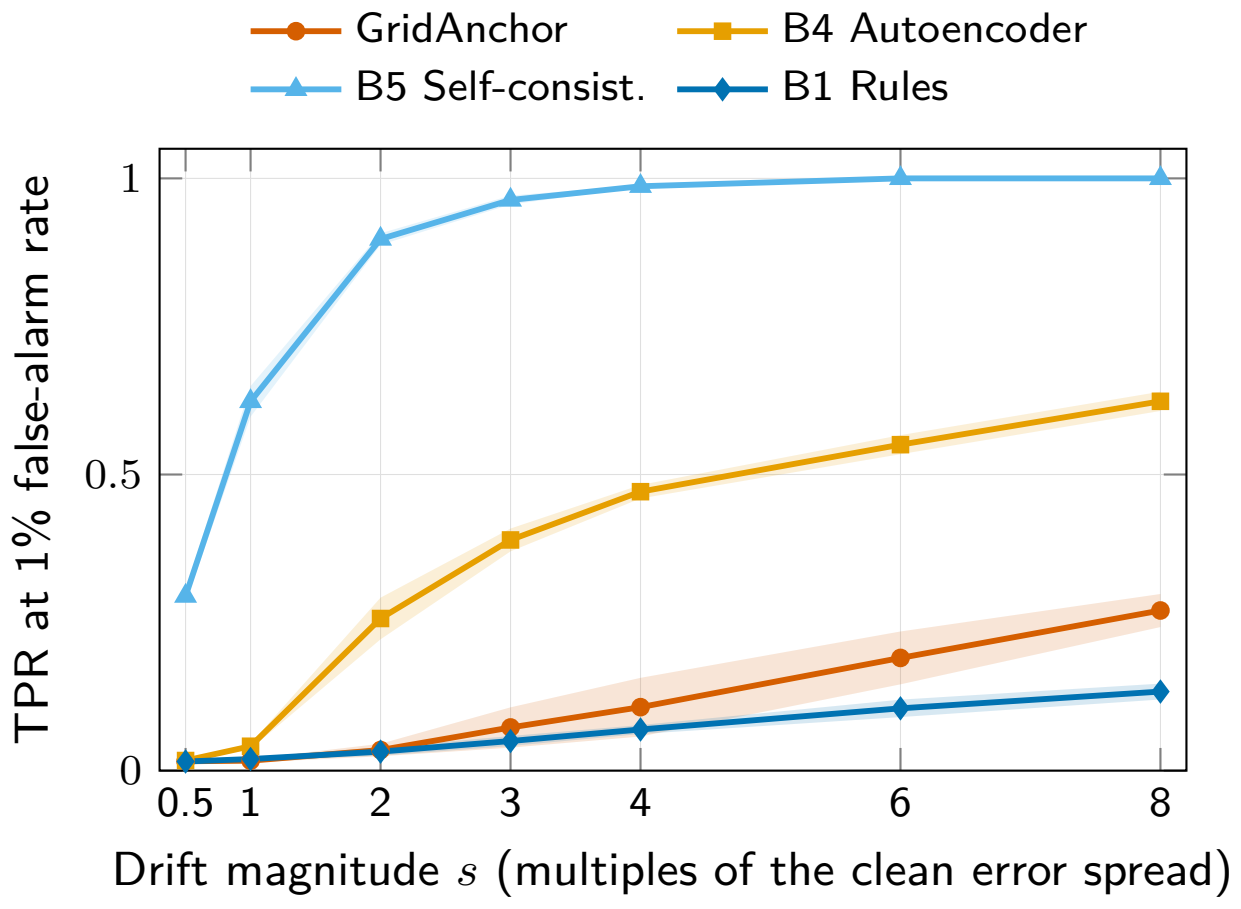

\subsection{Severity}
\label{sec:severity}
\Cref{fig:severity} varies the magnitude $s$ of drift episodes on F and E, in units of the clean error spread of each claim (mean over seeds, band for the standard deviation of GridAnchor).
GridAnchor detects these drifts poorly.
Its TPR at a FAR of 1\% rises from \resSweepTprOursLowPct\% at $s = 0.5$ to \resSweepTprOursHighPct\% at $s = \resSweepMax$, against \resSweepTprTopLowPct\% and \resSweepTprTopHighPct\% for the best of the three strongest baselines.
Yet even at $s = \resSweepMax$ only \resSweepCphShareHighPct\% of the drifted messages contain a claim beyond the tolerance: the PRGT is trained to leave the others unflagged, whereas B5 flags any deviation from its own recomputation.

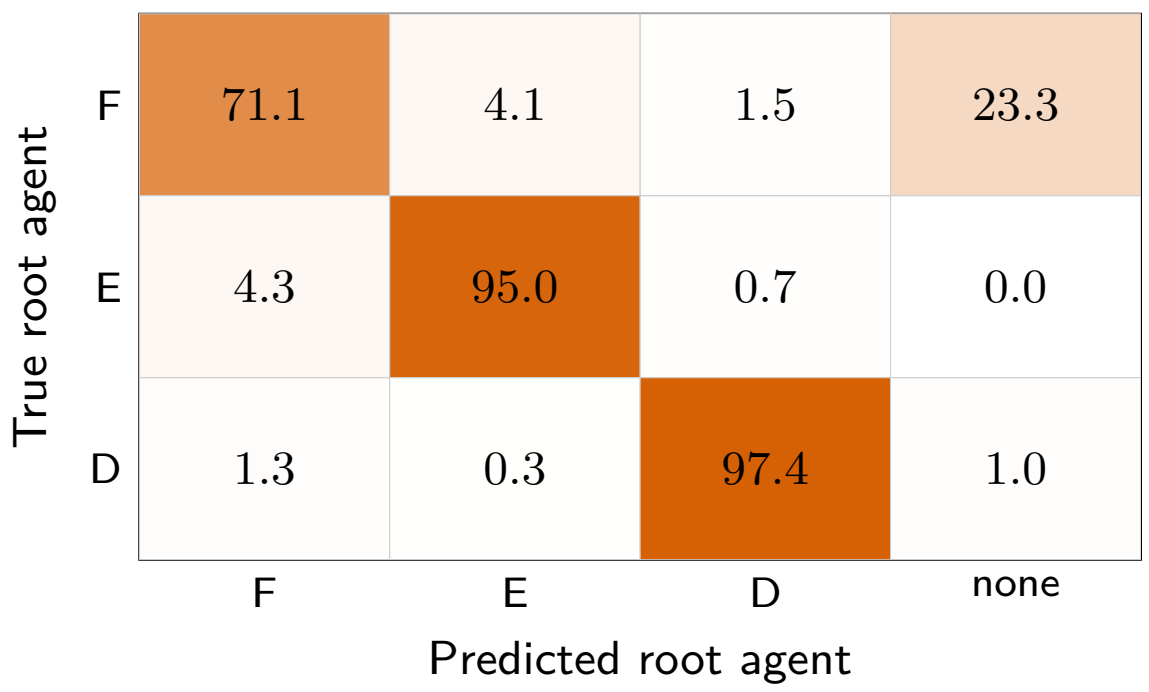
\begin{figure}[t]
\centering
\begin{tikzpicture}
\begin{axis}[paperstyle, width=0.98\columnwidth, height=0.62\columnwidth, enlargelimits=false, grid=none,
  colormap={ours}{color(0cm)=(white) color(1cm)=(cOurs)}, point meta min=0, point meta max=100,
  xmin=-0.5, xmax=3.5, ymin=-0.5, ymax=2.5, y dir=reverse,
  xtick={0,1,2,3}, xticklabels={F,E,D,none}, ytick={0,1,2}, yticklabels={F,E,D},
  tick style={draw=none}, xlabel={Predicted root agent}, ylabel={True root agent},
  nodes near coords={\pgfmathprintnumber[fixed, fixed zerofill, precision=1]{\pgfplotspointmeta}},
  nodes near coords style={font=\small, anchor=center, text=black}]
\addplot[matrix plot*, mesh/cols=4, point meta=explicit, draw=black!20, line width=0.2pt] table[x=x, y=y, meta=c, col sep=comma] {figures/data/fig6_confusion.csv};
\end{axis}
\end{tikzpicture}
\caption{Root-agent attribution by the root head.}
\label{fig:attribution}
\end{figure}

\subsection{Attribution and Cascades}
\label{sec:attribution}
On the steps with an injected episode, the root head names the correct agent in \resRootAccOursPct\% of the cases, and in \resRootAccDetOursPct\% of those in which a claim is flagged.
\Cref{fig:attribution} gives the row-normalized confusion matrix (seeds pooled).
The topological rule reaches \resRootAccTopoPct\% with the flags of GridAnchor and at most \resRootAccBestPct\% with those of a baseline (\resRootAccBestName).
A hallucination downstream also enlarges the residuals of the upstream claims to which the physics couples it.
The verifier flags the first hallucinated message of an episode after \resDelayOurs{} steps on average and misses \resDelayMissedOursPct\% of the episodes entirely, against \resDelayMissedBestPct\% for \resBestBaselineShort.

\begin{table*}[t]
\centering
\caption{Closed-loop operation on test days.}
\label{tab:closedloop}
\footnotesize
\begin{tabular}{@{}lccccccc@{}}
\toprule
 & Violation rate & Worst over- & Curtailed & Cost & SoC & Executed halluc. & Cascade con- \\
Method & (\% bus-steps) $\downarrow$ & voltage (p.u.) $\downarrow$ & (MWh) & (kEUR) & violations $\downarrow$ & actions $\downarrow$ & tainment (\%) $\uparrow$ \\
\midrule
Unguarded & 0.0700\std{0.0348} & 0.0145\std{0.0034} & 30.1\std{0.9} & 45.0\std{0.9} & 34.7\std{42.2} & 37.7\std{15.8} & 90.2\std{3.0} \\
B1 Rules & 0.0022\std{0.0032} & 0.0006\std{0.0008} & 38.7\std{5.4} & 49.9\std{3.6} & 0.0\std{0.0} & 0.0\std{0.0} & 100.0\std{0.0} \\
B2 WLS-BDD & 0.0002\std{0.0003} & 0.0001\std{0.0002} & 32.2\std{0.4} & 45.1\std{0.3} & 0.0\std{0.0} & 0.0\std{0.0} & 100.0\std{0.0} \\
B3 IsoForest & 0.0054\std{0.0058} & 0.0076\std{0.0068} & 33.9\std{3.4} & 47.9\std{3.0} & 0.3\std{0.5} & 9.7\std{2.4} & 96.0\std{1.2} \\
B4 Autoencoder & 0.0119\std{0.0112} & 0.0034\std{0.0029} & 38.8\std{2.8} & 49.3\std{3.8} & 4.3\std{6.1} & 6.3\std{6.9} & 98.5\std{1.1} \\
B5 Self-consist. & 0.0009\std{0.0013} & 0.0014\std{0.0019} & 40.1\std{6.4} & 48.2\std{1.3} & 0.0\std{0.0} & 0.3\std{0.5} & 100.0\std{0.0} \\
B6 Digital twin & 0.0002\std{0.0003} & 0.0001\std{0.0002} & 32.2\std{0.3} & 45.0\std{0.5} & 0.0\std{0.0} & 0.0\std{0.0} & 100.0\std{0.0} \\
B7 Safety filter & \best{0.0000}\std{0.0000} & \best{0.0000}\std{0.0000} & 50.0\std{3.0} & 52.6\std{0.6} & 0.0\std{0.0} & 0.0\std{0.0} & 100.0\std{0.0} \\
\midrule
GridAnchor (ours) & \best{0.0000}\std{0.0000} & \best{0.0000}\std{0.0000} & 32.7\std{0.8} & 45.1\std{0.4} & 0.0\std{0.0} & 0.0\std{0.0} & 100.0\std{0.0} \\
\midrule
Expert (reference) & 0.0000\std{0.0000} & 0.0000\std{0.0000} & 30.0\std{0.0} & 42.6\std{0.0} & 0.0\std{0.0} & 0.0\std{0.0} & 100.0\std{0.0} \\
\bottomrule
\end{tabular}

\end{table*}

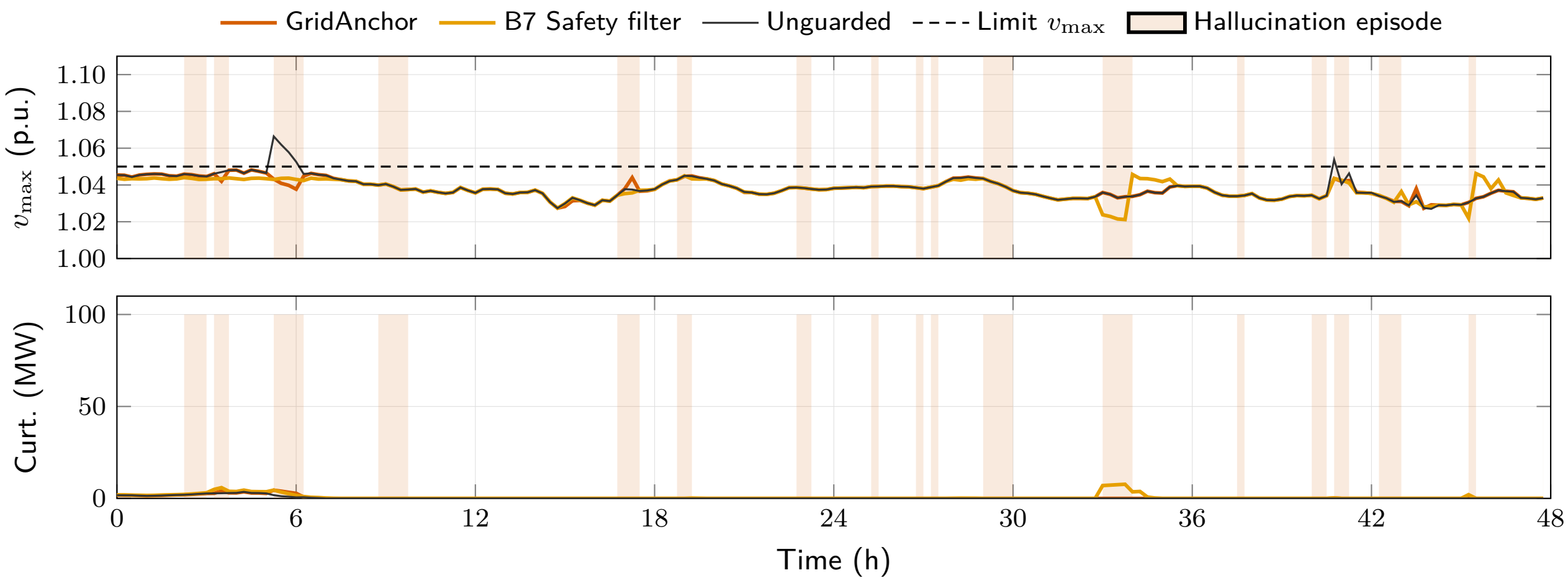
\begin{figure*}[t]
\centering
\begin{tikzpicture}
\begin{groupplot}[group style={group size=1 by 2, vertical sep=4mm, xlabels at=edge bottom, xticklabels at=edge bottom},
  paperstyle, width=0.97\textwidth, height=0.215\textwidth, xmin=0, xmax=48, xtick={0,6,...,48}]
\nextgroupplot[ymin=1.0, ymax=1.11, ytick={1.00,1.02,1.04,1.06,1.08,1.10}, ylabel={$v_{\max}$ (p.u.)},
  y tick label style={/pgf/number format/fixed, /pgf/number format/fixed zerofill, /pgf/number format/precision=2},
  legend columns=5, legend style={at={(0.5,1.04)}, anchor=south, /tikz/every even column/.append style={column sep=4pt}}]
\addplot[draw=none, fill=cOurs, fill opacity=0.13, const plot, forget plot] table[x=h, y expr=1.0+\thisrow{epi}*0.2, col sep=comma] {figures/data/fig7_case.csv} \closedcycle;
\addplot[cOurs, no marks] table[x=h, y=vours, col sep=comma] {figures/data/fig7_case.csv};
\addplot[\resCaseBaselineColor, no marks] table[x=h, y=vbaseline, col sep=comma] {figures/data/fig7_case.csv};
\addplot[black!80, no marks, line width=0.6pt] table[x=h, y=vunguarded, col sep=comma] {figures/data/fig7_case.csv};
\addplot[black, densely dashed, no marks, line width=0.6pt, domain=0:48, samples=2] {1.05};
\addlegendimage{area legend, fill=cOurs, fill opacity=0.13, draw=none}
\legend{GridAnchor, \resCaseBaselineName, Unguarded, Limit $v_{\max}$, Hallucination episode}
\nextgroupplot[ymin=0, ylabel={Curt. (MW)}, xlabel={Time (h)}]
\addplot[draw=none, fill=cOurs, fill opacity=0.13, const plot, forget plot] table[x=h, y expr=\thisrow{epi}*100, col sep=comma] {figures/data/fig7_case.csv} \closedcycle;
\addplot[cOurs, no marks] table[x=h, y=cours, col sep=comma] {figures/data/fig7_case.csv};
\addplot[\resCaseBaselineColor, no marks] table[x=h, y=cbaseline, col sep=comma] {figures/data/fig7_case.csv};
\addplot[black!80, no marks, line width=0.6pt] table[x=h, y=cunguarded, col sep=comma] {figures/data/fig7_case.csv};
\end{groupplot}
\end{tikzpicture}
\caption{Closed-loop case study on test day \resCaseDay.}
\label{fig:case}
\end{figure*}

\subsection{Closed-Loop Operation}
\label{sec:closedloop}
\Cref{tab:closedloop} reports the closed-loop runs on the \resClosedDays{} test days (mean and standard deviation over \resSeedsClosed{} seeds), with about \resEpisodesClosed{} episodes per seed over \resClosedSteps{} steps.
Containment is the share of episodes during which no executed action is a CPH.
The unguarded pipeline leaves \resViolUnguarded\% of the bus-steps outside the band and executes \resExecUnguarded{} hallucinated actions per run.
GridAnchor executes no hallucinated action and leaves no bus outside the band on any test day of any seed.
With the daily violation rate averaged over the seeds, the 95\% confidence interval of the unguarded pipeline is \resViolUnguardedCiLo{} to \resViolUnguardedCiHi\%.
The paired Wilcoxon test over the \resClosedStatDays{} days gives a corrected p-value of \resViolWilcoxonUnguardedP.

The detectors wrapped in the same mitigation sequence come close.
B2 and B6 leave \resViolBtwo\% and \resViolBsix\% of the bus-steps outside the band, B5 \resViolBfive\% and B1 \resViolBone\%.
B3 and B4 also let \resExecBthree{} and \resExecBfour{} hallucinated actions per run reach the plant.
None of these differences with GridAnchor is significant over \resClosedStatDays{} days (smallest corrected p-value \resViolWilcoxonMinBaseP, for \resViolWilcoxonMinBaseName).

GridAnchor curtails \resCurtOurs~MWh, \resCurtExtraPct\% more than the expert that knows the true state, against \resCurtExtraMinDetPct\% to \resCurtExtraMaxDetPct\% more for the six detector baselines.
It queries D again in \resRequeryPct\% of the steps and repairs the action in \resRepairPct\% of them.
The safety filter B7 also avoids every violation but curtails \resCurtBseven~MWh and cannot tell the operator which agent failed.
Repaired actions simulated again with pandapower differ from the plant simulator by at most $\resPpDevMax$~p.u.

The margin holds where the repair acts.
At the \resRepNOurs{} repaired steps of GridAnchor, $\delta(\bm{u}^\star) \le m_\beta$ in \resRepCovOursPct\% of the cases for a nominal \resRepCovNominalPct\%, and in \resRepCovFeOursPct\% of those during an episode on F or E.
The median error there, $\resRepErrMedianMpu \times 10^{-3}$~p.u., comes almost entirely from the grounded voltage estimate (median $\resRepEstMedianMpu \times 10^{-3}$~p.u.), not from the linear model ($\resRepLinMedianMpu \times 10^{-3}$~p.u.).
The offline margin of $\resMarginMpu \times 10^{-3}$~p.u., computed with clean estimates, therefore covered only \resRepCovOfflinePct\% of the repaired steps of the calibration runs.
The baselines ground flagged estimates by persistence but use the margin of GridAnchor, whose coverage at their repaired steps ranges from \resRepCovMinDetPct\% to \resRepCovMaxDetPct\%, so part of their residual violations may come from this mismatch.

\Cref{fig:case} shows the highest bus voltage and the curtailment over two test days for the unguarded pipeline, GridAnchor and \resCaseBaselineName, with the \resCaseEpisodes{} injected episodes shaded.
During these episodes the unguarded pipeline lets the highest voltage reach \resCaseVmaxUnguarded~p.u., against \resCaseVmaxOurs~p.u. with GridAnchor and \resCaseVmaxBaseline~p.u. with \resCaseBaselineName, which curtails more.
In the worst incident of the unguarded runs, a hallucination of type \resIncidentType{} on agent \resIncidentAgent{} drives \resIncidentBuses{} buses up to \resIncidentVUnguarded~p.u.; under GridAnchor the same step ends at \resIncidentVOurs~p.u.

\begin{table}[t]
\centering
\caption{Ablations of GridAnchor.}
\label{tab:ablation}
\footnotesize
\setlength{\tabcolsep}{3.4pt}
\begin{tabular}{@{}lcccc@{}}
\toprule
\multicolumn{5}{@{}l}{\textit{Detection (open loop, test days)}} \\
Variant & AUPRC $\uparrow$ & FAR & FDR & Root acc. $\uparrow$ \\
\midrule
GridAnchor (full) & 0.875 & 0.050 & 0.190 & 0.809 \\
\midrule
w/o residual tokens & 0.773 & 0.048 & 0.126 & 0.524 \\
w/o structural bias & 0.868 & 0.051 & 0.189 & 0.816 \\
w/o synthesis & 0.363 & 0.054 & 0.142 & 0.295 \\
Single conformal class & 0.865 & 0.050 & 0.177 & 0.811 \\
Per-claim $\alpha$, no BH & 0.876 & 0.054 & 0.984 & 0.892 \\
Topological attribution & 0.875 & 0.050 & 0.190 & 0.592 \\
\bottomrule
\end{tabular}

\vspace{2mm}
\begin{tabular}{@{}lccc@{}}
\toprule
\multicolumn{4}{@{}l}{\textit{Mitigation (closed loop)}} \\
Variant & Viol. (\%) $\downarrow$ & Curt. (MWh) & Exec. $\downarrow$ \\
\midrule
GridAnchor (full) & 0.0000 & 33.6 & 0.0 \\
\midrule
F grounded by reconstruction & 0.0000 & 32.2 & 0.0 \\
No re-query & 0.0000 & 33.4 & 0.0 \\
No margin ($m{=}0$) & 0.0141 & 31.3 & 0.0 \\
\bottomrule
\end{tabular}

\end{table}

\subsection{Ablations and Validation Loop}
\label{sec:ablation}
\Cref{tab:ablation} removes one element of the layer at a time, all rows on the same seed (FAR at $\alpha = 0.05$, FDR at $q = 0.1$).
Without the residual tokens, the PRGT reads only the claims and the raw measurements and its AUPRC falls from \resAuprcFullAbl{} to \resAuprcNoRes.
Without taxonomy-driven synthesis, the claim head sees only clean targets, because natural hallucinations carry no weight in its loss, and the AUPRC falls to \resAuprcNoSynth.
Its scores still rank claims, since the reconstruction and physics losses shape the representation and the head lowers the scores of clean inputs only.
The root head of this variant learns only the class none, so its attribution reduces to the pattern of flags (accuracy \resRootHeadNoSynth, against \resRootTopoNoSynth{} for the topological rule on the same flags).
The structural bias has a small effect (\resAuprcNoBias), possibly because the residuals already encode the electrical coupling.

A single conformal class keeps the AUPRC (\resAuprcSingle) and the pooled FAR (\resFarSingleFivePct\% for a nominal 5\%).
The FAR of the six claim classes then ranges from \resFarClassMinSinglePct\% to \resFarClassMaxSinglePct\%, however, against \resFarClassMinOursPct\% to \resFarClassMaxOursPct\% with the Mondrian classes.
Per-claim tests at level $\alpha = q$ without the BH step raise the FDR to \resFdrNoBh, and the topological attribution rule lowers the root accuracy from \resRootHeadAbl{} to \resRootTopoAbl.

In the mitigation block, the margin is the component that prevents violations on this seed: without it, \resViolNoMargin\% of the bus-steps leave the band.
Grounding the forecasts by their reconstructions also avoids every violation and curtails \resCurtAltGround~MWh instead of \resCurtFullAbl~MWh, and without the second query of D the curtailment is \resCurtNoRequery~MWh.
On the closed-loop validation days, the selected rule had given \resValViolHybrid\% of violations, against \resValViolRec\% with reconstructions for both agents and \resValViolUnguarded\% without protection.
Grounding both agents by persistence needs a margin of $\resMarginClPersistMpu \times 10^{-3}$~p.u., because an outdated voltage estimate deviates far from the current state.
Its repairs then curtail \resValCurtPersist~MWh on the validation days, against \resValCurtExpert~MWh for the expert.

The configuration of the PRGT results from a logged loop on validation days with \resLoopTried{} single-lever changes, of which \resLoopKept{} were kept, raising the validation AUPRC from \resValAuprcBase{} to \resValAuprcFrozen.
Two of the three targets set before the loop were met on validation days: calibration, and a violation rate \resValViolRedPct\% below that of the unguarded pipeline for \resValCurtExtraPct\% more curtailment than the expert.
The detection target, a margin of 0.05 in AUPRC over the strongest baseline, was missed (\resValMargin{} over \resValBestName).

\begin{table*}[t]
\centering
\caption{Transfer, robustness and latency.}
\label{tab:generalization}
\footnotesize
\setlength{\tabcolsep}{3.2pt}
\begin{minipage}[t]{0.47\textwidth}
\centering
\begin{tabular}[t]{@{}lccc@{}}
\toprule
\multicolumn{4}{@{}l}{\textit{(a) Transfer to the 122-bus feeder}} \\
Detector & AUPRC $\uparrow$ & FAR & FDR \\
\midrule
GridAnchor, in-domain & 0.788 & 0.051 & 0.199 \\
GridAnchor, zero-shot & 0.719\std{0.025} & 0.050\std{0.002} & 0.176\std{0.007} \\
\quad source calibration & 0.698\std{0.081} & 0.060\std{0.015} & 0.212\std{0.028} \\
\midrule
B1 Rules & 0.661\std{0.025} & 0.054\std{0.001} & 0.126\std{0.001} \\
B2 WLS-BDD & 0.586\std{0.047} & 0.051\std{0.001} & 0.097\std{0.007} \\
B3 IsoForest & 0.385\std{0.022} & 0.045\std{0.001} & 0.068\std{0.004} \\
B4 Autoencoder & 0.740\std{0.030} & 0.051\std{0.000} & 0.179\std{0.009} \\
B5 Self-consist. & 0.783\std{0.023} & 0.047\std{0.000} & 0.183\std{0.002} \\
B6 Digital twin & 0.458\std{0.088} & 0.021\std{0.000} & 0.042\std{0.002} \\
\midrule
\multicolumn{4}{@{}l}{\textit{(b) Extra missing meter readings}} \\
Missing & AUPRC $\uparrow$ & FAR & FDR \\
\midrule
none & 0.846\std{0.033} & 0.051\std{0.001} & 0.181\std{0.007} \\
5\,\% & 0.847\std{0.032} & 0.050\std{0.001} & 0.180\std{0.006} \\
10\,\% & 0.847\std{0.032} & 0.050\std{0.001} & 0.178\std{0.005} \\
20\,\% & 0.847\std{0.032} & 0.049\std{0.002} & 0.178\std{0.006} \\
\bottomrule
\end{tabular}
\end{minipage}\hfill
\begin{minipage}[t]{0.5\textwidth}
\centering
\setlength{\tabcolsep}{2.6pt}
\begin{tabular}[t]{@{}lcccccccc@{}}
\toprule
\multicolumn{9}{@{}l}{\textit{(c) Per-type AUPRC, type withheld from training}} \\
Training & H1 & H2 & H3 & H4 & H5 & H6 & H7 & H8 \\
\midrule
All types & 0.73 & 0.85 & 0.98 & 0.17 & 0.89 & 0.97 & 0.90 & 0.56 \\
Withheld & 0.70 & 0.87 & 0.99 & 0.04 & 0.91 & 0.94 & 0.24 & 0.56 \\
\bottomrule
\end{tabular}

\vspace{3mm}
\setlength{\tabcolsep}{2.6pt}
\begin{tabular}[t]{@{}lccccccc@{}}
\toprule
\multicolumn{8}{@{}l}{\textit{(d) Latency of one step (ms, one CPU thread)}} \\
 & & Resi- & & Conf. & & \multicolumn{2}{c}{Total} \\
Feeder & Buses & duals & PRGT & + BH & Repair & median & p95 \\
\midrule
LV-rural1 & 15 & 0.6 & 1.1 & 0.3 & 1.9 & 3.8 & 4.9 \\
LV-semiurb4 & 44 & 0.6 & 1.5 & 0.4 & 2.4 & 5.0 & 7.5 \\
MV-rural & 99 & 0.7 & 2.9 & 0.4 & 5.2 & 9.3 & 10.0 \\
MV-semiurb & 122 & 0.7 & 2.8 & 0.4 & 6.3 & 10.1 & 15.2 \\
MV-urban & 144 & 1.1 & 6.1 & 0.6 & 8.0 & 16.4 & 18.7 \\
\bottomrule
\end{tabular}
\end{minipage}

\end{table*}

\subsection{Generalization, Robustness and Latency}
\label{sec:generalization}
\Cref{tab:generalization} collects four studies of GridAnchor.
In the first, the PRGT trained on the main feeder is applied to the 122-bus feeder with its weights unchanged.
Only the residual scales, the subspaces and the calibration set are recomputed from clean logs of that feeder.
It reaches an AUPRC of \resTransferAuprcZero, against \resTransferAuprcIn{} for a PRGT trained on that feeder (one seed).
Its FAR at a nominal 5\% is \resTransferFarRecalPct\%, and \resTransferFarNoRecalPct\% with the calibration set of the source feeder.
The baselines are fitted on the clean log of the target feeder with the settings selected on the main feeder.
Of these, \resTransferNBaseAboveZero{} exceed the zero-shot verifier, up to \resTransferAuprcBestBase{} for \resTransferBestBaseName, so zero-shot transfer avoids retraining at a cost in accuracy.

Withholding one type from training changes the AUPRC on that type by less than 0.05 for \resLotoNSmallDrop{} of the eight types (median drop \resLotoMedianDrop, one seed).
Drifts and stale replays are the exceptions: without them in training, the AUPRC falls from \resLotoSeenHSeven{} to \resLotoUnseenHSeven{} on H7 and from \resLotoSeenHFour{} to \resLotoUnseenHFour{} on H4.
Their residuals have the size of the clean errors, so the verifier must learn their temporal signature from examples.
Even then, the AUPRC of the main models on H7 ranges from \resTypeHSevenMin{} to \resTypeHSevenMax{} across seeds.

With 20\% of extra missing meter readings, the AUPRC is \resMissingAuprcMax{} and the FAR \resMissingFarMaxPct\% at a nominal 5\%, almost the values without them, consistent with residual families that mostly do not use the meters.
One verification step of GridAnchor, repair program included, takes a median of \resLatMedianMain~ms and a 95th percentile of \resLatPninetyfiveMain~ms on the main feeder on one CPU thread.
The 95th percentile is \resLatStepShare\% of the control step and stays below \resLatMaxFeeder~ms on all five feeders.
The calibrated decision alone takes \resLatOurs~ms for GridAnchor and \resLatBfive~ms for B5, which runs five forecasters, five estimators and five decoding passes of the SLM.

\section{Discussion}
\label{sec:discussion}
Physics residuals help where a hallucination contradicts a quantity that the plant constrains.
Voltage estimates, actions and their rationales are tied to the measurements by the power flow, and GridAnchor detects them best (AUPRC \resAuprcEOurs{} on E and \resAuprcDOurs{} on D).
Redundancy is the stronger test for a forecast, since the load of a single bus is barely visible in the measurements of the same step (\resAuprcFBfive{} for B5 against \resAuprcFOurs{} on F).
Knowledge of the tolerances matters as much as the architecture.
GridAnchor learns from its labels where a change within the tolerance ends and a hallucination begins, and given the same tolerances the self-consistency check overtakes it in the primary metric.

The failure cases follow from the same argument.
A drift below the noise floor leaves residuals of the size of the clean errors and is detected only when drifts are part of the training data.
A coherent stale replay leaves little residual (AUPRC \resTypeHFourOurs{} on H4).
Exchangeability fails inside an episode, where the FDR reaches several times $q$.
It also fails at the repaired steps for a margin computed from clean estimates, which covered \resRepCovOfflinePct\% of the errors there because the grounded voltage estimate, not the linear model, dominates them.
Calibrated at the repaired steps of closed-loop runs, the margin holds on the test days, at the price of a tightening of $\resMarginClMpu \times 10^{-3}$ instead of $\resMarginMpu \times 10^{-3}$~p.u. at the few repaired steps.

Among the negative findings, the validation loop missed its detection target, and the difference with the strongest of the six baselines is within the confidence interval.
In closed loop, the bad-data and twin-based detectors with the same mitigation left only \resViolBtwo\% of the bus-steps outside the band with slightly less curtailment, and no difference with GridAnchor is significant over \resClosedDays{} days.
The structural attention bias contributed little on a feeder of \resBusTotal{} buses, and on the transfer feeder two refitted baselines detected better than the zero-shot verifier.

The study has limitations that bound its conclusions.
All results come from simulation on two SimBench feeders, with the plant model also used to label the data, and no field measurement or hardware was involved.
The hallucinations are synthetic, and a deployed large language model may fail differently; the dispatcher solves a one-step problem and is a small language model with \resDParamsM~million parameters trained from scratch.
The scale is modest, with \resSeeds{} seeds, \resClosedDays{} closed-loop test days and one seed for the secondary studies, and the margin assumes that the calibration days resemble the operating days.

For deployment, the results suggest combining physics residuals for estimates and actions with tolerance-aware redundancy for forecasts in one calibrated layer.
Both calibrations, of the p-values and of the repair margin, need recent clean logs and must be repeated after any change of topology, agents, meters or grounding rule.
Alarms are more reliable for messages than for claims, so grounding should be validated on many more days before it acts on a plant.

\section{Conclusion}
\label{sec:conclusion}
We formalized cyber-physical hallucinations of specialized learned agents, built a benchmark of eight hallucination types on SimBench feeders and proposed GridAnchor, a verification layer that scores, calibrates, attributes and repairs.
On the main feeder, GridAnchor reaches a message-level AUPRC of \resAuprcOurs, on par with the strongest of six baselines (\resBestBaselineShort, \resAuprcBestBaseline) and significantly better than \resWilcoxonNSigBase{} of them.
A self-consistency check that knows the label tolerances reaches \resAuprcBfiveTol, but GridAnchor leads on estimates and actions.
It keeps the false-alarm rate within \resFarMaxGapPp{} percentage points of the nominal level on steps without an episode and names the root agent in \resRootAccOursPct\% of the hallucinated steps.
In closed loop it leaves no bus outside the voltage band on the test days, with \resCurtExtraPct\% more curtailment than the expert, and simpler detectors wrapped in the same grounded re-query and repair come close.
The study also marks the limits of physical grounding.
Errors that leave no trace in concurrent measurements, such as a forecast error on one bus or a stale but coherent message, call for redundancy or longer temporal context.
Drifts must be part of the training data.
Collateral flags inside an episode weaken the false-discovery guarantee, and the repair margin holds only when calibrated on the errors of the grounded estimates.
Future work will combine physics residuals with tolerance-aware redundancy in one calibrated test and use field data.

\section*{Author Contributions}
\textbf{Abderaouf Bahi}: Writing – original draft, Software, Validation, Methodology, Investigation, Conceptualization. 
\textbf{Hasan Masrur}: Writing – review and editing, Resources, Supervision.

\section*{Acknowledgment}
Not Applicable.

\section*{Ethical Approval} 
Not Applicable.

\section*{Conflict of Interest}
The authors declare that they have no known competing financial interests or personal relationships that could have appeared to influence the work reported in this paper.

\section*{Funding}
This research received no external funding.

\section*{Data Availability}
We have shared the GitHub repository with data and code.

\section*{Declaration of Generative AI and AI-assisted technologies in the writing process}
The authors used Claude (Opus 5.5) to assist with drafting and language refinement.
All outputs were reviewed and edited by the authors, who take full responsibility
for the final content.

\appendix
\section{Proofs}
\label{sec:proofs}

\begin{pf}[Proof of \cref{prop:far}]
By exchangeability, the rank of $s(c)$ among the $n_k + 1$ scores, counted from the largest with ties against the test claim, is stochastically at least uniform on $\{1, \dots, n_k+1\}$, and \eqref{eq:pvalue} is this rank divided by $n_k+1$.
Hence $\Pr\{p(c) \le \alpha\} \le \lfloor \alpha (n_k+1) \rfloor / (n_k+1) \le \alpha$, with equality to $\lfloor \alpha (n_k+1) \rfloor / (n_k+1) \ge \alpha - 1/(n_k+1)$ when the scores are almost surely distinct.
This is the standard rank argument of conformal prediction \cite{vovk2022algorithmic,angelopoulos2023conformal}.
\end{pf}

\begin{pf}[Proof of \cref{prop:fdr}]
Within one class, the p-values of the test claims share the same calibration scores and are positively regression dependent on the subset of clean claims (PRDS) \cite{bates2023testing}.
The calibration sets of different classes are disjoint and the test claims are independent given the calibration data, so the classes form independent blocks, and a vector of independent PRDS blocks is PRDS on the union of the clean subsets.
The BH procedure at level $q$ on $m$ PRDS p-values, $m_0$ of which belong to clean claims, has a false discovery rate of at most $q\,m_0/m$ \cite{benjamini1995controlling,bates2023testing}.
\end{pf}

\begin{pf}[Proof of \cref{prop:repair}]
Let $\bm{u}^\star$ be the solution of \eqref{eq:repair} with $\bm{s} = \bm{0}$, and let $\delta = \delta(\bm{u}^\star)$ be the end-to-end error \eqref{eq:delta} at the test step.
The margin $m_\beta$ is the $\lceil (n+1)(1-\beta) \rceil$-th smallest of $n$ calibration errors, so exchangeability gives $\Pr\{\delta \le m_\beta\} \ge 1-\beta$ by the argument of \cref{prop:far}.
On this event, the constraints of \eqref{eq:repair} with zero slack imply $v_{\min} \le v_n(\bm{u}^\star) \le v_{\max}$ for every bus $n$.
For the second claim, the feasible set of \eqref{eq:repair} with $\bm{s} = \bm{0}$ is a convex polyhedron and the objective is strictly convex in $\bm{u}$ because $\bm{W}$ is positive definite, so $\bm{u}^\star$ is the unique feasible action closest to $\bm{u}_{\mathrm{D}}$ in the $\bm{W}$-norm.
\end{pf}

The exchangeability required by \cref{prop:repair} refers to the action actually evaluated and to the estimate that enters the program, which a margin computed from clean estimates and unrepaired actions does not satisfy.

\section{Hyperparameters and Tokenizer Grammar}
\label{sec:hyper}
The PRGT uses $d = 64$, $L = 3$, $H = 4$, a feed-forward width of $4d$, dropout 0.1 and 16 distance buckets.
It is trained with AdamW (learning rate $10^{-3}$ with cosine decay, weight decay $10^{-4}$, batches of 32 steps) for \resPrgtEpochsMax{} epochs of 150 batches.
The focal parameter is 2 and the loss weights are $\lambda_{\mathrm{rec}} = 0.5$, $\lambda_{\mathrm{phys}} = 0.1$ and $\lambda_{\mathrm{root}} = 0.5$, with one injected realization of the training days.

The previous grounded claims seen in training are the clean previous claims, replaced by the received ones with probability 0.3.
During the memory curriculum of the last five epochs, they are replaced by self-generated memory with probability 0.5.
The cross-sectional subspaces of the main feeder have \resSubDimLoad, \resSubDimRes{} and \resSubDimVolt{} dimensions for the load, renewable and voltage claims.
The SLM is trained with AdamW (learning rate $3 \times 10^{-4}$, batches of 64 sequences) on completion tokens only.

The vocabulary has 31 symbols (digits, signs, time, zone and field tags, a separator and control symbols), and a prompt holds the quarter-hour index and the zone values as fixed-width integers.
A completion must match the grammar ``(B sign digit$^4$ K digit$^2$)$^Z$ X digit$^4$ C digit$^5$ end''.
Any other token sequence is a format hallucination, in which case the previous action is held and the repair starts from it.

\bibliographystyle{cas-model2-names-unsrt}
\bibliography{refs}

\end{document}